\documentclass[a4paper,11pt,times,astrosym]{aastex631}   

\usepackage[english]{babel}
\usepackage{amsmath}
\usepackage{graphicx}
\usepackage{natbib}

\submitjournal{ApJS}

\shorttitle{Science Opportunities for IMAP-Lo Observations of Interstellar Neutral H and D} 

\newcommand{\kms}{~km~s$^{-1}$}

\newcommand{\cc}{~cm$^{-3}$}
\newcommand{\cmsq}{~cm$^{-2}$}
\newcommand{\persec}{~s$^{-1}$}

\newcommand{\hpri}{\ensuremath{\text{H}_{\text{pri}}}}
\newcommand{\hsec}{\ensuremath{\text{H}_{\text{sec}}}}
\newcommand{\hmix}{\ensuremath{\text{H}_{\text{mix}}}}
\newcommand{\hsum}{\ensuremath{\text{H}_{\text{sum}}}}

\newcommand{\dpri}{\ensuremath{\text{D}_{\text{pri}}}}
\newcommand{\dsec}{\ensuremath{\text{D}_{\text{sec}}}}

\newcommand{\cD}{\ensuremath{c_\text{D,ISN}}}
\newcommand{\cDterr}{\ensuremath{c_{\text{D,terr}}}}

\begin{document}

\title{Science Opportunities for IMAP-Lo Observations of Interstellar Neutral Hydrogen and Deuterium During a Maximum of Solar Activity}

\correspondingauthor{M.A. Kubiak}
\email{mkubiak@cbk.waw.pl} 

\author[0000-0002-5204-9645]{M.A. Kubiak}
\affil{Space Research Centre PAS (CBK PAN), Bartycka 18a, 00-716 Warsaw, Poland}

\author[0000-0003-3957-2359]{M. Bzowski}
\affil{Space Research Centre PAS (CBK PAN), Bartycka 18a, 00-716 Warsaw, Poland}

\author[0000-0002-2745-6978]{E. M{\"o}bius}
\affil{University of New Hampshire, Durham, NH}

\author[0000-0002-3737-9283]{N.A. Schwadron}
\affil{University of New Hampshire, Durham, NH}

\begin{abstract}
Direct-sampling observations of interstellar neutral gas, including hydrogen and deuterium, have been performed for more than one cycle of solar activity by IBEX. IBEX  viewing is restricted to directions perpendicular to the spacecraft--Sun line, which  limits the observations to several months each year. This restriction is removed in a forthcoming mission Interstellar Mapping and Acceleration Probe. The IMAP-Lo instrument will have a capability of adjusting the angle of its boresight with the spacecraft rotation axis. We continue a series of studies of resulting science opportunities. We adopt a schedule of adjusting the boresight angle suggested by \citet{kubiak_etal:23a} and focus on interstellar hydrogen and deuterium during solar maximum epoch. Based on extensive set of simulations, we identify the times during calendar year and elongation angles of the boresight needed to measure the abundance of D/H at the termination shock and unambiguously observe interstellar H without contribution from interstellar He. Furthermore, IMAP-Lo will be able to resolve the primary and secondary populations, in particular to view the secondary population with little contribution from the primary. We show that the expected signal is sensitive to details of radiation pressure, particularly its dependence on radial speed of the atoms, and to details of the behavior of the distribution function of the primary and secondary populations at the heliopause. Therefore, IMAP-Lo will be able to provide observations needed to address compelling questions of the heliospheric physics, and even general astrophysics.
\end{abstract}


\section{Introduction}
\label{sec:intro}
\noindent
The heliosphere is formed and shaped by an interaction between the solar wind plasma and partly ionized interstellar matter. The main component of interstellar matter is hydrogen. 
In addition to H, He, and other species, interstellar matter has a small fraction of a hydrogen isotope deuterium. 

Since interstellar neutral (ISN) atoms penetrate freely inside the heliopause, the information on the physical state and processes within the immediate solar neighborhood in the Galaxy can be retrieved by direct sampling observations of various species of ISN atoms at 1 au from the Sun, such as those performed by Interstellar Boundary Explorer \citep[IBEX, ][]{mccomas_etal:09c, fuselier_etal:09b, mobius_etal:09a, mobius_etal:09b} since 2009, and in a near future by Interstellar Mapping and Acceleration Probe \citep[IMAP, ][]{mccomas_etal:18b}.

Deuterium is important cosmologically because the entire population of D atoms is believed to have been created during Big Bang nucleosynthesis, as there are no known astrophysical processes that produce D atoms in significant amounts inside of stars \citep{linsky_etal:06a, prantzos:96a, prantzos:07a}.
However, processes are operating in stellar cores that transform D into heavier species, thus likely reducing the D/H ratio in galaxies over times. 
Consequently, the local abundance of D is a tracer of Big Bang nucleosynthesis and the chemical evolution of matter in our galaxy. 
The study of ISN D is one of the science goals of the IMAP mission.

Retrieval and interpretation of information carried by ISN H and D atoms is challenging. Ionization due to interaction with the solar wind and solar EUV radiation depletes ISN H and D atoms between the heliopause and the detection sites at 1 au. 
Additionally, the solar resonant radiation pressure counteracts solar gravity and as a result, ISN H and D atoms have much slower speeds at 1 au than ISN He, for which radiation pressure is negligible \citep{bzowski_etal:97, tarnopolski_bzowski:08a, IKL:23a}.
This reduces the impact energy of ISN H and D atoms and consequently their detection efficiency.
Additionally, radiation pressure is radial velocity-dependent due to the Doppler effect \citep{tarnopolski_bzowski:09, IKL:18b} for both ISN H and D, and the strength of all these effects varies with the phase of solar cycle. 
All this  makes the analysis even more challenging and results in the necessity to use appropriate atom and ionization tracking procedures.

In addition to the modifications of the ISN atom flow inside the heliopause, ISN atoms strongly interact with the disturbed interstellar plasma ahead of the heliopause and create the so-called secondary populations of ISN gas. 
The existence of the secondary population of ISN H has been postulated based on modeling insight \citep[e.g.,][]{baranov_malama:93}. 
The secondary population of ISN He was discovered based on direct-sampling observations from IBEX \citep{bzowski_etal:12a, kubiak_etal:14a, bzowski_etal:17a}.
In the case of ISN H, its presence is known from indirect evidence.
\citet{lallement_etal:05a} measured the direction and magnitude of the inflow velocity of ISN H based on spectroscopic observations of the heliospheric resonant backscatter glow (the helioglow) performed at 1 au. 
They found the inflow velocity vector significantly different to that obtained from direct-sampling observations of ISN He performed by Ulysees \citep{witte:04}. 
The existence of such a difference had been postulated based on kinetic modeling of the heliosphere \citep{baranov_etal:98a}. 
While modification of ISN He due to the interaction in the outer heliosheath is known to be weak \citep{swaczyna_etal:23a}, that of ISN H is strong \citep{izmodenov_etal:03b}. 
As a result, the dominant population of ISN H at the heliopause is the secondary. 
The primary is significantly depleted, and its flow parameters modified, as suggested by \citet{bzowski_etal:08a} based on interpretation of observations of hydrogen pickup ions on Ulysses.
However, the aforementioned observations of the helioglow, and, to our knowledge, any other available heliospheric observations are not able to unambiguously resolve the primary and secondary ISN H.

Since the signal due to ISN H is to a large extent swamped in that due to ISN He \citep{saul_etal:12a, rahmanifard_etal:19a} due to the detection technique of ISN atoms, distinguishing of the two populations might be feasible using a specially devised geometry of direct sampling observations.
This distinction would be welcome because it would hopefully facilitate studying the processes operating in the outer heliosheath and modifications of the flow of the primary ISN H in this region, and consequently, the physical state of ISN H in the pristine interstellar matter.

The cosmic abundance of ISN D relative to H is approximately equal to $1.6 \times 10^{-5}$ \citep{hebrard_etal:99a, linsky_etal:06a}. 
Consequently, the absolute flux of ISN D at 1 au is much lower than that of ISN H.

Measurement aspects of ISN D on IBEX were presented by \citet{tarnopolski_bzowski:08a} and \citet{kubiak_etal:13a}. 
These latter authors suggested that detection of just several individual ISN D atoms during one year of IBEX operation during low solar activity can be expected.
And sure enough, based on this insight,  \citet{rodriguez_etal:13a} and \citet{rodriguez_etal:14a} found evidence for ISN D in the IBEX-Lo data, albeit based
on less than 10 atoms, along with many more D atoms sputtered off the terrestrial water layer covering the IBEX-Lo conversion surface.
Clearly, acquiring a number of ISN D atoms on IMAP that would permit a statistical analysis with a reasonable confidence requires identifying best suitable times and observation geometries during the mission. 

This paper presents such a study, as well as a study aimed at optimizing future observations of ISN H by IMAP-Lo.
It is the most recent item in a line of papers discussing science opportunities given by the IMAP-Lo experiment, and particularly by its ability to change the viewing direction in flight \citep{mccomas_etal:18b}.
\citet{sokol_etal:19c} presented the science opportunities and thoroughly discussed measurement and observation aspects of ISN H, He, Ne, and O, as well as D, including the expected flux magnitudes of these species during solar minimum and maximum activity conditions. 
They also pointed out the solar elongation angles of the boresight of the IMAP-Lo instrument during the year needed to follow the peaks of the ISN species.
\citet{schwadron_etal:22a} and \citet{bzowski_etal:22a} discussed the reasons of and suggested observation geometry options best suitable for removing the inflow parameter correlations obtained in analysis of IBEX-Lo observations.
\citet{bzowski_etal:23a} suggested a calibration-free method to derive the ionization rate of ISN He by comparison observations of the direct and indirect beams, which may be a viable way to address a hypothesis by \citet{swaczyna_etal:22b} that the ionization rate of this species is substantially biased.
\citet[\citet{kubiak_etal:23a}]{kubiak_etal:23a} suggested a two-year scenario of setting the elongation angle of the IMAP-Lo instrument allowing to exercise all of the science opportunities suggested by \citet{sokol_etal:19c} during two calendar years of operations and focused on presenting various aspects of such observations of the heavy species, including the primary and secondary populations of ISN He and O, and the primary population of ISN Ne.

In this paper, we focus on IMAP-Lo observations of ISN H and D. 
We adopt the pivot angle adjustment scenario suggested by \citet{kubiak_etal:23a}.
For H (Section \ref{sec:H}), we point out the observation conditions for which ISN H can be observed with a minimum contribution from ISN He to the signal (Section \ref{sec:onlyH}). We also suggest how to resolve the primary and secondary populations.
Section \ref{sec:onlyHpr} discusses possible observations of the primary population of ISN H (\hpri) with as little contribution from the secondary as possible, and 
Section \ref{sec:onlyHsc} presents the opposite: the times and observation geometries best suitable for viewing the secondary H (\hsec) with a minimum contribution from \hpri.
Since the flow parameters of the H populations are known much less precisely than those of ISN He, in Section \ref{sec:Hshift} we discuss the sensitivity of the measured flux to the flow parameters.
The challenges with unambiguous separation of the two ISN H populations are presented in Section \ref{sec:Hmix}.
Finally, we investigate the sensitivity of the expected signal to details of radiation pressure, which based on IBEX-Lo analyses \citep{schwadron_etal:13a,rahmanifard_etal:19a,katushkina_etal:21b} was suggested to differ from that obtained from analysis of the solar spectral observations (\citet{lemaire_etal:15a}, \citet{tarnopolski_bzowski:09}, \citet{IKL:18a}, \citet{IKL:20a}; Section \ref{sec:VrH}).
Prospects for detection of ISN D and prerequisites for maximization of the count statistics are discussed in Section \ref{sec:D}. We compare the fluxes of D (Section \ref{sec:Dflux}), count rate of ISN D and terrestial D sputtered from ISN He (Section \ref{sec:crD}) and ratio of the latter count rates (Section \ref{sec:cRatio}).
The paper is concluded in Section \ref{sec:conc}.

\section{Simulations and the elongation angle schedule}
\label{sec:simu}
\noindent
We adopt the same simulation pool that was used by \citet{sokol_etal:19c}, \citet{bzowski_etal:22a}, \citet{bzowski_etal:23a}, and \citet{kubiak_etal:23a}, appropriately extended with simulations used to investigate the sensitivity of the signal to different inflow parameters, for comparison of the two-population model with a one-population approximation, and the model with the dependence of the radiation pressure used in the atom tracking switched off (Sections \ref{sec:Hshift}-\ref{sec:VrH}). 

The simulations  yield the total flux in atoms \cmsq \persec{}, performed using the WTPM model (\citet{tarnopolski_bzowski:09} and \citet{tarnopolski_bzowski:08a}) for ISN H and D, respectively, with details concerning detection on IBEX and IMAP given in \citet{sokol_etal:15b} and \citet{sokol_etal:19c}, on a 2D grid in the space of day of the year (DOY) and elongation angle $\varepsilon$. Details of the grid and time coverage are presented in \citet{sokol_etal:19c}. 
The presented fluxes are filtered by the collimator, but no other interactions with the instrument are simulated.

In the simulations, we used the same ionization rate and radiation pressure models as those employed in \citet{sokol_etal:19c}, i.e., by \citet{sokol_etal:19a} and \citet{IKL:18a}, respectively. 
Even though since the publication of the original paper by \citet{sokol_etal:19c} both of these models have been updated, we decided to stick to their older versions in the new simulation set performed for this paper to maintain homogeneity and cross-comparability of the results. 
The ionization rate model is heliolatitude- and time-dependent, and includes charge exchange, photoionization, and electron-impact reactions. 
The ionization rates for H and D are adopted as identical. 
The time series used are presented in Figure 3 in \citet{sokol_etal:19a}, and their heliolatitude dependence in Figure 5 in the same paper.

\begin{figure}
\centering
\includegraphics[width=0.45\textwidth]{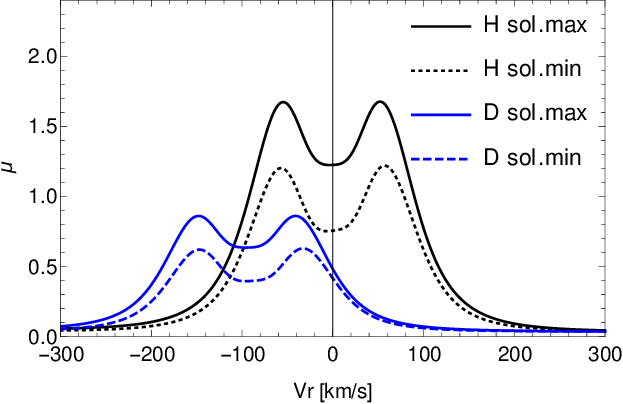}
\includegraphics[width=0.51\textwidth]{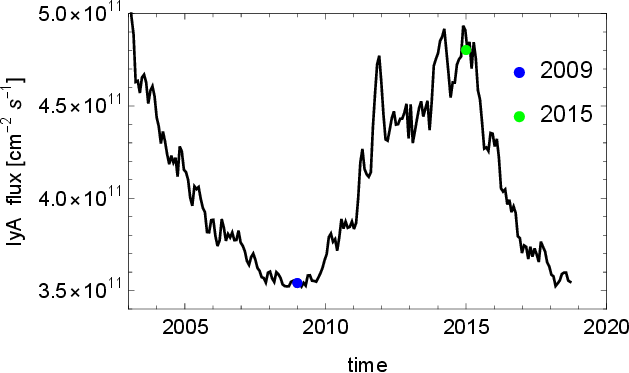}
\caption{
Solar gravity compensation factor $\mu(v_r)$ for H (black) and D (blue) for the epochs of solar maximum (solid lines) and minimum conditions (broken lines) in 2015.0 and 2009.0,  respectively, based on \citet{IKL:18a}, are shown in the left panel. The model of the evolution of the $\mu$ factor is driven by the magnitude of the total flux in the solar Lyman-$\alpha$ line, shown in the right panel. The epochs selected to represent the solar maximum and minimum conditions are marked with the green and blue dots. }
\label{fig:HDRadPress}
\end{figure}
 
Radiation pressure was adopted after \citet{IKL:18a}. 
The magnitudes of the solar gravity compensation factor due to radiation pressure $\mu(v_r)$ for H and D for the epochs 2015.0 and 2009.0 are displayed in the left panel of Figure \ref{fig:HDRadPress}. 
The differences between the radiation pressure acting in H and D atoms is caused on the one hand by the isotope effect, which shifts the profile characteristic for D leftward in the figure by 81.3802 \kms, which results from the difference between the H and D Lyman-$\alpha$ wavelengths of $-0.0333$ nm, and on the other hand by the difference in masses between H and D atoms, which is responsible for reduction of the compensation factor for D approximately by half  (precisely, it is a factor of 0.50038436).

The magnitude of the radiation pressure force depends on radial velocities of individual atoms and drops with square of solar distance \citep[however, see discussion of opacity effects in ][]{IKL:22a}. 
The radial velocity dependence follows the spectral flux profile of the solar Lyman-$\alpha$ line.
The absolute magnitude of the force also scales with the magnitude of the solar Lyman-$\alpha$ flux, adopted from \citet{machol_etal:19a}. 
This quantity is presented in the right panel of Figure \ref{fig:HDRadPress}.
In the calculations, the dependence on the Lyman-$\alpha$ flux corresponds effectively to the dependence of the force on time, and together with the dependence on $v_r$ is accounted for in the simulation module responsible for numerical calculation of the trajectories of individual atoms.

The $\mu$ factor is adopted as independent of the distance to the Sun. 
Strictly speaking, this assumption is true only in the optically thin approximation. 
The validity of this approximation for the flux of ISN H observed at 1 au was demonstrated by \citet{IKL:22a} and we use it in the present paper. 

In this paper, we use the same simulation pool as that used by \citet{sokol_etal:19c} for the primary and secondary populations of ISN H and D. 
Additionally, for this paper we performed simulations for ISN H discussed in Section \ref{sec:Hshift} for the primary and secondary populations for varying, but coordinated inflow directions. 

Another set of simulations, also performed on the full grid of (DOY, $\varepsilon$) parameters and discussed in Section \ref{sec:Hmix}, uses one population of ISN H with the inflow parameters being appropriate weighted averages of the parameters of the primary and secondary populations. 

Still another set of simulations, again on the full grid of (DOY, $\varepsilon$), assumes that the solar spectral flux in the Lyman-$\alpha$ line is flat, i.e., there is no dependence of radiation pressure on radial velocities of the atoms. 
Results of comparison of this model with the more realistic one are presented in Section \ref{sec:VrH}. 

Most of the the simulations used in this paper were performed assuming realistic time variations of the ionization rates. 
The time dependence of the ionization and solar radiation factors is an inherent feature of the WTPM model. 
Since IMAP will be launched during solar maximum conditions, we decided -- following \citet{kubiak_etal:23a} -- to focus on solar maximum conditions and consequently used the simulations for the days of the 2015 year. An exception is discussion of ISN D, where we additionally use simulations performed for solar minimum conditions, specifically for days during the 2009 year.  

We stress that the simulations presented in the paper must not be regarded as IMAP-Lo signal predictions. 
It is not possible to know in advance the magnitudes of the solar output responsible for ionization of ISN H and D and the radiation pressure at the time of future observations. 
In particular, the strength of the forthcoming maximum of the solar cycle is expected to be greater than that of the last maximum, which would likely reduce the magnitudes of the fluxes. 
Moreover, the models of the ionization rate and radiation pressure undergo permanent refinement. 
\citet{sokol_etal:19c} used the model of the ionization rate by \citet{sokol_etal:19a} and the model of radiation pressure by \citet{IKL:18a}. 
Since that time, one iteration of the radiation pressure model \citep{IKL:20a} and two iterations of the ionization rate model \citep{sokol_etal:20a, porowski_etal:22a}, with additional refinement by \citet{porowski_etal:23a}, have been published.
From the viewpoint of magnitudes of relevant quantities, these are incremental modifications, and not qualitative changes. 

To qualify for discussion, the flux of ISN H had to exceed a threshold of 10 atoms \cmsq \persec{} and an energy threshold of 10~eV. 
The magnitudes of these thresholds result from analysis of IBEX-Lo observations.

The schedule of the elongation angle settings proposed for the two years of the nominal IMAP mission is presented by \citet{kubiak_etal:23a}, see their Figure 5 and Table 1. 
We urge the reader to consult them as the elongation angle schedule adopted by these authors is essential for our paper as well. 
In brief, we propose different elongation angle settings for the first and the second year of operations. 
The $\varepsilon$ settings make carefully selected horizontal lines in the (DOY, $\varepsilon$) space.
The $\varepsilon$ settings are organized in 4-day cycles (A--D), with the days A and C optimized for ISN gas observations, and B and D for ENA observations. 
The energy settings of the instrument are identical for all days throughout the two years covered by the schedule. 
With this, observations taken during the ENA days can potentially be used for ISN analysis and vice versa. 

The baseline velocities and temperatures of \hpri{} and \hsec{} used in the simulations are identical to those used by \citet{sokol_etal:19c}.
They are listed in the first row of Table \ref{tab:shiftParam}.
The densities are adopted from \citet{bzowski_etal:08a} and \citet{bzowski_etal:09a}. 
They were obtained mostly on analysis of pickup ions observed by Ulysses.
We regard them as conservative estimates: $n_{\hpri} = 3.1\times 10^{-2}$ \cc{} and $n_{\hsec} = 5.4189 \times 10^{-2}$ \cc.
For \hmix, we adopt the sum of these two values.
For ISN D, we used identical parameters to those used for ISN H, only the densities were scaled down by the D/H abundance factor characteristic for the local interstellar medium, adopted at 15.6 ppm after \citet{linsky_etal:06a}: $n_{\dpri} = 4.836 \times 10^{-7}$ \cc, $n_{\dsec} = 8.453484 \times 10^{-7}$ \cc.

However, \citet{swaczyna_etal:20a} used observations of pickup ions performed by the New Horizons mission much farther away from the Sun, in a region outside of the cavity in the ISN H density carved by the solar factors, and obtained the total density of ISN H larger by $\sim 50$\%.
If this is true, then the total magnitude of ISN H at 1 au might be larger by about 50\% because the flux is linearly proportional to the density. However, a breakdown of ISN H into \hpri{} and \hsec{} under these conditions would require a considerable re-evaluation, which is outside the scope of this paper. 

The simulated fluxes of all ISN species and their respective populations for the $\varepsilon$ setting scenario used in our paper, including \hpri{} and \hsec{}, is presented in the electronic figures discussed in the Appendix in \citet{kubiak_etal:23a}. 
An exception is ISN D, which requires a special processing of observations due to a very low magnitude of its flux at the detector. 
This issue is discussed in Section \ref{sec:D}.

\section{Hydrogen}
\label{sec:H}
\noindent
Hydrogen is the most abundant species in the Local Interstellar Medium, but because of the joint action of the solar resonant radiation pressure and ionization processes its flux at 1 au is reduced by at least two orders of magnitude and becomes lower than that of ISN He. 
Its flow is substantially modified in the outer heliosheath \citep{lallement_bertaux:90a, lallement_etal:93a} due to charge-exchange and collision processing \citep{baranov_malama:93, baranov_izmodenov:06a, swaczyna_etal:21a, rahmanifard_etal:23a}.

Modeling studies of the processes operating in the OHS resulted in the conclusion that the ISN H gas inside the heliopause can be represented as a superposition of two non-interacting distribution functions, corresponding to the primary and secondary populations \citep[e.g.][]{izmodenov:01, katushkina_izmodenov:10}. 
This seems feasible because ISN H inside the heliopause is a collisionless gas, with the atoms following trajectories governed solely by the solar gravity and radiation pressure forces.
In the past, the simplest possible approximation of these functions was frequently used:
it was assumed that both the primary and the secondary populations at a certain reference distance from the Sun, typically around 150 au or just inside the heliopause, are represented by the Maxwell-Boltzmann function with different densities, bulk velocity vectors and temperatures for the two populations \citep[e.g.,][]{bzowski_etal:08a}. 
This assumption is also used in this paper.
However, up to now these two distribution functions have not been resolved observationally.

While average parameters of the flow have been determined from observations of the helioglow \citep{quemerais_etal:99, lallement_etal:05a, lallement_etal:10a,katushkina_etal:15a, koutroumpa_etal:17a, bzowski_etal:23b}, the parameters of the two individual populations could only be obtained indirectly and depend on the adopted model of the heliospheric boundary region. 

Based on analysis of pickup ions measured by Ulysses and modeling of the heliospheric boundary region \citep{izmodenov_etal:03b}, \citet{bzowski_etal:08a} suggested the flow parameters of the primary and secondary populations of ISN H at the termination shock conforming with the observations of pickup ions. 
The primary population was obtained a little cooled and accelerated relative to the unperturbed ISN H, but the secondary was heated by a factor of 2.5, slowed by a factor of two, and had a density almost twice larger. 
The parent model of the heliosphere had no magnetic field, and the flow was axially symmetric. 
However, the distortion of the heliosphere from axial symmetry was postulated based on modeling \citep{ratkiewicz_etal:02a} and confirmed by observations of heliospheric energetic neutral atoms \citep{zirnstein_etal:16c,mccomas_etal:20a, schwadron_mccomas:21a,zirnstein_etal:22b}, anisotropies in the cosmic ray flux \citep{schwadron_etal:14a}, sounding of the heliopause distance by ENA response to solar-cycle variations of the solar wind \citep{reisenfeld_etal:21a}, and by the fact that the mean velocity vectors of ISN H and ISN He inside the heliosphere differ from each other.
Consequently, the inflow directions of the primary and secondary populations are expected to be drawn aside along the plane determined by the vectors of Sun's velocity across the Local Interstellar Medium and interstellar magnetic field. 
This latter insight is obtained based on analysis of the IBEX Ribbon by \citet{zirnstein_etal:16b} and independent determinations of the inflow parameters of the primary \citep{bzowski_etal:15a, schwadron_etal:15a} and secondary populations of ISN He \citep{kubiak_etal:16a}. 

Analysis of ISN H observations on IBEX-Lo has been challenging because of the detection technology used. 
ISN H is observed by means of H$^-$ ions formed due to a capture of an electron from a special conversion surface by an impacting ISN H atom. 
However, ISN He is also observed due to H$^-$ ions, which are sputtered from the conversion surfce by the impacting ISN He atoms \citep{wurz_etal:08a}.
Thus, a separation of the signals from these two interstellar species has only been possible by meticulous analysis of times of flight of the ions registered by the instrument \citep[see, e.g.,][]{park_etal:16a}.
In addition, since the impact energy of ISN H is only about 10 eV, and the sensitivity of the instrument decreases with decreasing energies, the detection efficiency for ISN H is the lowest \citep{fuselier_etal:09b}.

In this paper, we look for the regions in the (DOY, $\varepsilon$) space where the conditions for measurements of ISN H will be the best: the flux of He ISN will be the smallest, and differences in various assumptions will be the most visible in the flux as a function of spin angle.


\subsection{H without He}
\label{sec:onlyH}
\begin{figure}
\centering
\includegraphics[width=0.7\textwidth]{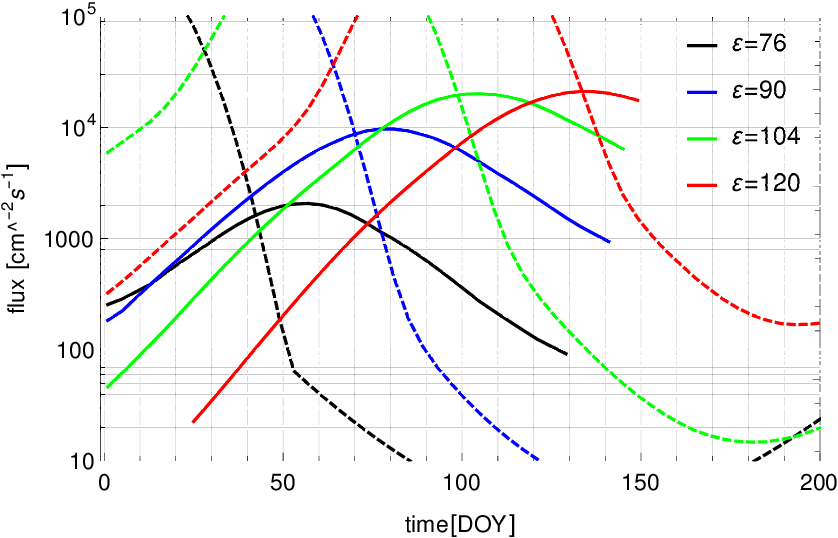}
\caption{
Magnitudes of the total flux of ISN H (a sum of the primary and secondary populations), marked with solid lines, compared with those for ISN He (broken lines) as a function of DOY for selected elongation angles $\varepsilon$ (color-coded). 
The end of the solid lines at the right side is defined by the energy threshold criterion: the impact energy drops below 10 eV. 
The beginning of the intervals for the ISN H campaign is defined as the DOYs for which the solid line becomes larger than the broken line in the corresponding color.
}
\label{fig:onlyHflux}
\end{figure}

\noindent
In this section, we are looking for opportunities to observe ISN H with the contribution from ISN He minimized.
Such locations along the Earth orbit for the IBEX-Lo observation geometry (i.e., for $\varepsilon = 90\degr$) have been identified by several authors \citep{saul_etal:12a, saul_etal:13a, rahmanifard_etal:19a, galli_etal:19a}.
During the ENA days B, D during the first year, and during days B during the second year IMAP-Lo observes at the identical elongation angle as IBEX-Lo does.
With this, for the DOYs 73---137 we have an opportunity to view ISN H without contribution from He. 
The overall sensitivity of IMAP-Lo is substantially increased over that of IBEX-Lo because it features a larger geometric factor by about x3.6 \citep{mccomas_etal:18b}, reduced background by more than one order of magnitude, and increased foreground-free viewing time from the L1 vantage point. In addition, the mounting on the pivot platform allows optimized viewing of ISN H over substantially extended times over the year.
The magnitudes of the maximum flux of ISN H compared with those of ISN He for the elongation $\varepsilon = 90\degr$ is drawn with green lines in Figure \ref{fig:onlyHflux}.

However, there are more regions in the (DOY, $\varepsilon$) space where pure ISN H can be observed. 
\citet{kubiak_etal:23a} proposed a Hydrogen Campaign for IMAP-Lo, marked in their paper in Figure 5 with gray dots.
During the first year of observations, for $\varepsilon = 76\degr$ the ISN H signal with little contribution from ISN He will be gathered during DOYs 45---130, during DOYs 73---137 for $\varepsilon = 90\degr$, and for $\varepsilon = 104\degr$ during DOYs 100---145.
During the second year of operations, pure H observations for $\varepsilon = 90\degr$ will be repeated, and in addition we expect a pure H signal for $\varepsilon = 120\degr$.
The beginnings of these intervals are selected so that the maximum of the ISN H signal is greater than that of ISN He, as shown for the aforementioned $\varepsilon$ angles in Figure \ref{fig:onlyHflux}. 
The end of the campaign for individual elongations coincide with the impact energy dropping below the detection threshold, adopted here at 10 eV. 

The selected regions in the (DOY, $\varepsilon$) space are advantageous for several reasons: there is very little or no ISN He expected, there are ranges in the spin angle where either solely \hsec{}, or \hpri{} with little contribution from \hsec{} are expected. 
Consequently, for a given DOY and $\varepsilon$ we can analyze both of the populations, and additionally investigate the response of the instrument to ISN H without contamination by H$^-$ ions sputtered by ISN He. 

It is interesting to compare the relations between the magnitudes of the He and H fluxes for different elongations shown in Figure \ref{fig:onlyHflux}. 
For $\varepsilon = 90\degr$ (IBEX-like), the ISN H flux becomes dominant very shortly before it attains its yearly maximum. 
This means that a half of the yearly beam is obscured by ISN He.
For $\varepsilon = 76\degr$, the ISN H flux exceeds that of ISN He earlier than in the former case, but the absolute magnitude of the flux is lower. 
It seems thus that the most advantageous is the elongation $\varepsilon = 104\degr$, which offers a high flux for ISN H that emerges from the signal due to ISN He sufficiently early. 
Concerning $\varepsilon = 120\degr$, the drop of the impact energy below the detection threshold prevents a measurement of a high flux for the DOYs when the signal from ISN He is low.

\begin{figure}
\centering
\includegraphics[width=0.24\textwidth]{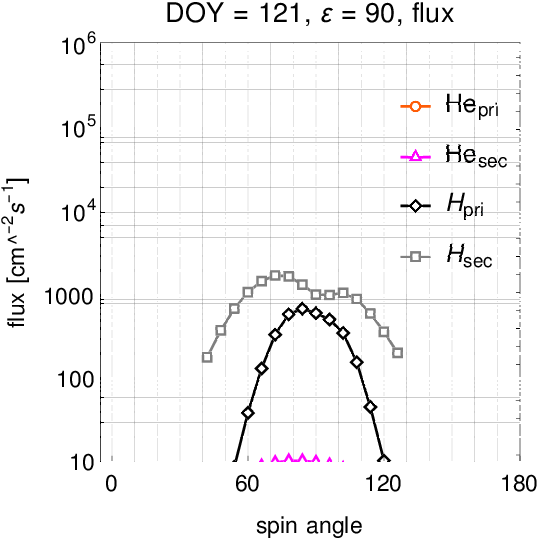}
\includegraphics[width=0.24\textwidth]{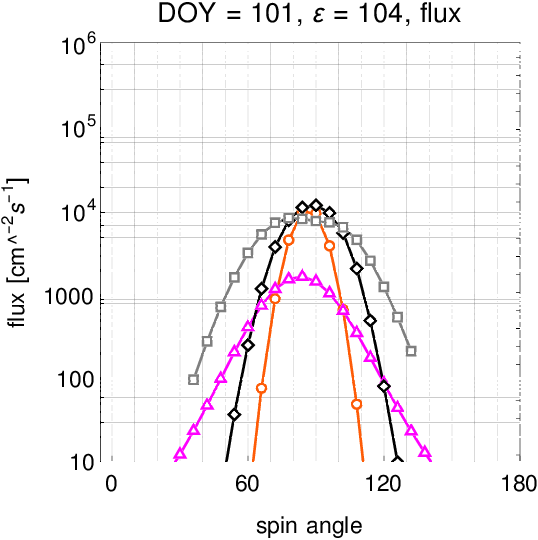}
\includegraphics[width=0.24\textwidth]{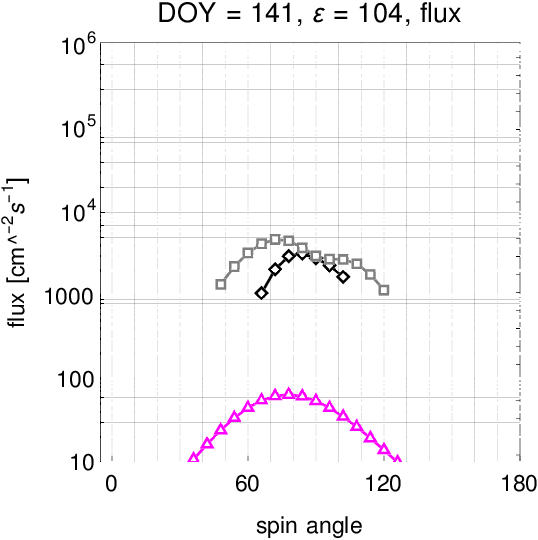}
\includegraphics[width=0.24\textwidth]{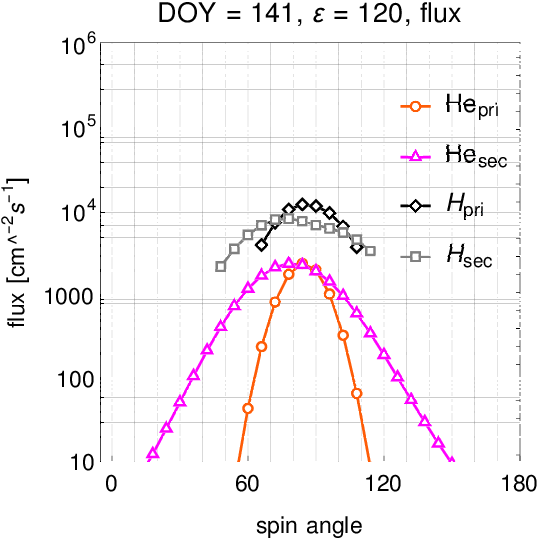}
\caption{
Example (DOY, $\varepsilon$) combinations illustrating the issue of separation of the signal due to ISN H from that due to ISN He.
The panels present the flux vs spin angle for selected (DOY, $\varepsilon$) combinations.
The species and populations are color-coded. See text for discussion. 
}
\label{fig:onlyH}
\end{figure}
\begin{figure}
\centering
\includegraphics[width=0.33\textwidth]{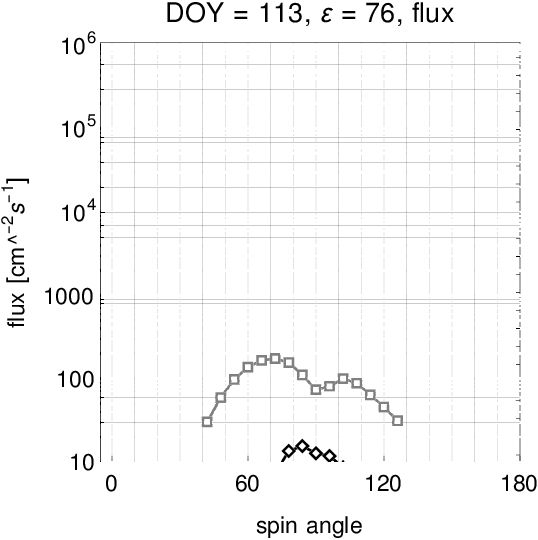}
\includegraphics[width=0.33\textwidth]{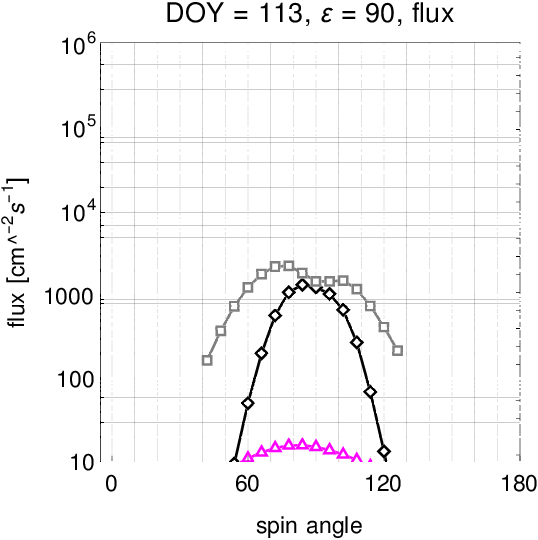}
\includegraphics[width=0.33\textwidth]{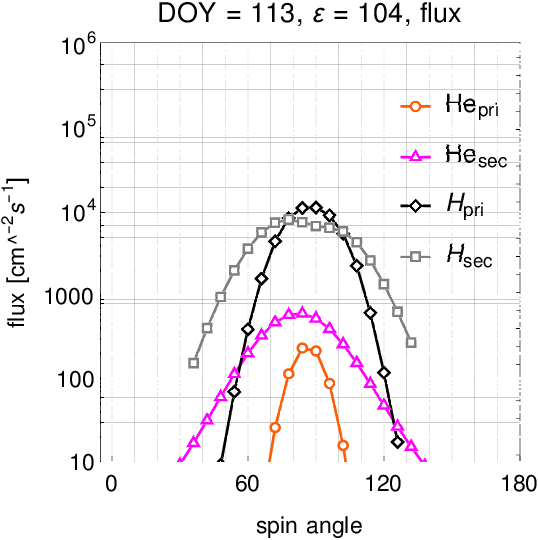}

\caption{ Illustration of an opportunity to scan the ISN H beam simultaneously in DOY and $\varepsilon$ cuts. 
This opportunity opens on DOY 109 and closes on DOY 117. The expected fluxes of \hpri{} and \hsec{} are illustrated in the three panels for the elongations $\varepsilon = 76\degr, 90\degr$, and 104\degr, respectively, from left to right, for DOY 113, which belongs to this interval. 
}
\label{fig:onlyHEpsScan}
\end{figure}

This discussion is illustrated by examples shown in Figure \ref{fig:onlyH}. 
The IBEX-like geometry illustrates the first panel, for DOY 121. The primary ISN He is absent, the secondary He is barely visible, likely below the detection threshold. 
The signal is dominated by ISN H.
The secondary population exceeds the primary, only for a few pixels near the peak the difference is relatively small.

In the second and third panels, we present the other preferred elongation for ISN H, for $\varepsilon = 104\degr$. 
For DOY 101 (second panel), the signals from ISN He and H are close to each other for a few pixels near the peak, but for all the other ones ISN H dominates, and the absolute magnitude of the flux is close to the global maximum.
40 days later, (third panel), there is very little of the secondary He and no primary He, only the primary and secondary H remain.  
The fourth panel presents a different elongation to the latter one ($\varepsilon = 120\degr$, not 104\degr) for the same DOY.
Clearly, the flux from ISN H is much larger than that from ISN He, but the energy (not shown) is so low that only a few ISN H pixels exceed the energy cutoff. 

Another interesting opportunity for analysis of ISN H opens on DOY 100, and closes on DOY 130 during the first year of IMAP operations. 
During this interval, for all three elongations suggested for the first year of operations, i.e., 76\degr, 90\degr, and 104\degr, the signal is expected to comprise almost solely ISN H, as shown for an example DOY 113 in Figure \ref{fig:onlyHEpsScan}. 
Consequently, we can scan across the beam of ISN H both in DOY and in the elongation. 

In all, the interval when ISN H dominates for at least one $\varepsilon$ angle within the four-day cycle of observations spans DOYs 45---145. 
This provides about 100 DOYs, i.e., long enough to hope to break the correlation between the parameters of inflow of ISN H, expected in direct-sampling measurements performed for vantage locations distributed along short orbital arc around the Sun \citep{bzowski_etal:22a}. 

\subsection{Viewing individual populations}
\label{sec:only1popu}
\noindent
Identification of the viewing conditions for which individual populations, \hpri{} or \hsec, are visible without a significant contributions from the other one would be welcome to facilitate determination of their inflow parameters. 
When the observed beam comprises two populations in comparable amounts, one can either determine an estimate of the mean inflow parameters or to fit two sets of inflow parameters, i.e., the components of the velocity, the temperatures, and the ratio of the densities in the source region, which makes a total of 9 parameters. 
This is computationally demanding, and in this section we verify if can be avoided.

\subsubsection{\hpri{} without \hsec}
\label{sec:onlyHpr}
\noindent
Our analysis showed that observation of \hpri{} without significant contribution from \hsec{} is challenging.
This is because inside the heliopause, the secondary population is more abundant than the primary by  a factor of 2.5 \citep{bzowski_etal:08a}. 
On the other hand, the temperature of \hpri{} is much lower than that of \hsec, because the beam of primary atoms able to penetrate sunward to 1 au is more collimated than that of the hotter secondary population.
Similarly, as for He, the beam of \hpri{} is much narrower than that for \hsec, and therefore there is no observation geometry guaranteeing viewing \hpri{} with no \hsec.
Nevertheless, we have identified some regions in the (DOY, $\varepsilon$) space conforming within the adopted observation scenario where $\hpri > \hsec$. Some examples are shown in Figure~\ref{fig:onlyH}, panels 2 and 4, and in the third panel of Figure \ref{fig:onlyHsc}. 
But the excess of \hpri{} over \hsec{} is not large, so we do not expect that it will be possible to separate \hpri{} from \hsec{} clearly and without modeling support. 

\subsubsection{\hsec{} without \hpri}
\label{sec:onlyHsc}
\noindent
Conversely, opportunities to view \hsec{} with a small contribution from \hpri{} are plentiful. 

For $\varepsilon = 76\degr$, this opportunity begins about DOY 85 and ends on DOY 117. 
For $\varepsilon = 90\degr$, the time interval where the signal contains almost solely \hsec{} comprises DOYs 121---157. 
For $\varepsilon = 104\degr$, this interval is much shorter: DOYs 145--165. 

The observations for $\varepsilon = 90\degr$ are suggested to be performed for two days during each four-day cycle during the first year, and for one day per cycle during the second year.
These cases are the most valuable from the viewpoint of analysis of ISN H, and in particular of \hsec.
The flow parameters obtained from these time intervals correspond solely to the secondary population. 
Thus, it is possible to determine the flow parameters for pure \hsec{}. 
Due to a relatively short orbital arc, breaking the parameter correlation will likely not be possible. 
Some alleviation of this drawback may be possible using the observations for $\varepsilon = 76\degr$  and $\varepsilon = 104\degr$ during the first year. 
With this, there is an interval of $\sim 100$ days with observations of pure \hsec{} separated in time, which might help reduce the parameter correlation.

\begin{figure}[!ht]
\centering
\includegraphics[width=0.24\textwidth]{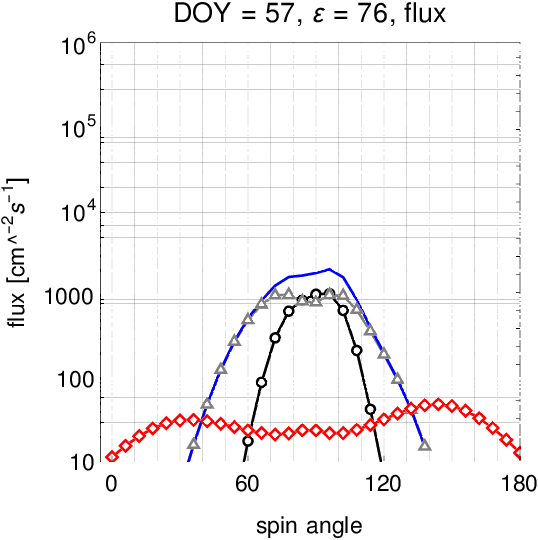}
\includegraphics[width=0.24\textwidth]{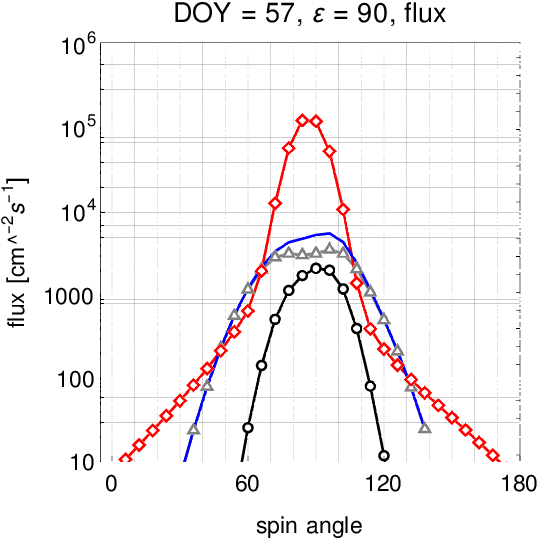}
\includegraphics[width=0.24\textwidth]{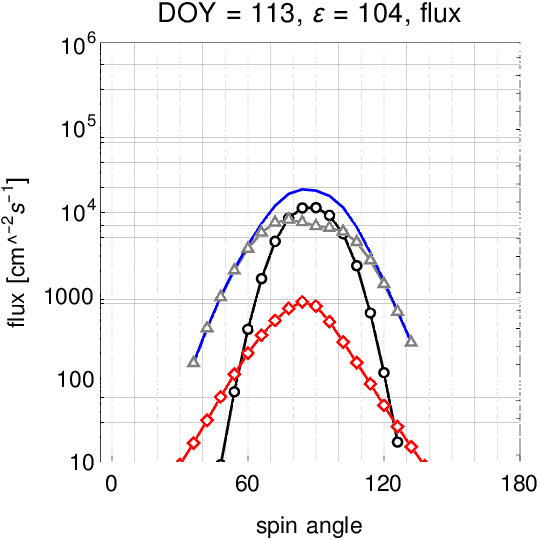}
\includegraphics[width=0.24\textwidth]{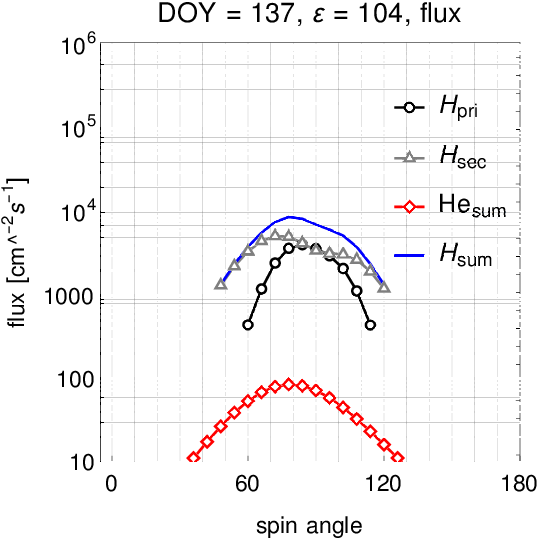}
\caption{
Example combinations of (DOY, $\varepsilon$) well suitable for observation of \hsec, contrasted with a case which is not well suitable, presented in the second panel from the left (where we have a lot contribution from He). 
We show \hpri, \hsec, and their sum as a function of spin angle, color-coded as indicated in the panels. 
Additionally, a sum of the primary and secondary He is presented. 
See text for details. 
}
\label{fig:onlyHsc}
\end{figure}

Illustrative examples are shown in Figure \ref{fig:onlyHsc}, where in addition to \hpri{} and \hsec{} (drawn in black and gray, respectively), we also show a sum of the primary and secondary populations of ISN He, drawn in red. 
The two leftmost panels present the flux for DOY 57. 
For $\varepsilon = 76\degr$, a vast majority of the signal is due to H.
In the wings of the distribution, \hsec{} is by far dominant.
For $\varepsilon = 90\degr$, ISN H is swamped by ISN He. 
The third panel represents DOY 113, $\varepsilon = 104\degr$; the signal is composed of a combination of \hpri{} and \hsec{} and it is not dominated by ISN He. 
\hsec{} dominates at the wings.
The rightmost panel represents the same elongation, but  a later DOY 137, when the impact energy of ISN H drops so much that even though the magnitude of the flux is large, the signal is expected to be cut off at the wings because of the energy threshold.
Thus, the usability of this combination depends on the energy sensitivity of the instrument, even though ISN He is not expected to obscure ISN H. 

An interesting aspect of the ISN population viewing is the double-hump feature, visible, in Figures \ref{fig:onlyH}---\ref{fig:onlyHsc}. We verified by modeling that it is due to ionization, and appears more readily for the flows with lower Mach numbers, like those for the secondary population. 
Reduction in the ionization rate results in gradual filling of the trough between the humps. 
Thus, observation of the hump suggests that the flow being observed is mostly due to \hsec. 
Note, however, that in some cases the double-hump feature may be absent in the total flux that is being measured, because the measurement does not differentiate between the primary and secondary populations -- the instrument sees their sum, and the double-hump appears for the slower and warmer \hsec, but not in the cooler and faster \hpri. 
In such situations, as can be seen, e.g., in the third panel of Figure \ref{fig:onlyHsc}, the wings of the observed flux profile correspond to \hsec, which would have featured a double-hump structure in the absence of \hpri, but since this latter population is present, the double-hump is not visible because it is filled by \hpri. 

Details of the viewing conditions for \hpri{} and \hsec{} can be obtained consulting the sequence of electronic figures in the Appendix in \citet{kubiak_etal:23a}.

\subsection{Sensitivity of the hydrogen flux to variations in the inflow parameters}
\label{sec:Hshift}
\noindent
IMAP-Lo seems to have a potential to resolve \hpri{} and \hsec.
While the fluxes of these populations are largely co-located in the spin angle space, it seems feasible to separate them with some modeling support, especially that the results are only weakly sensitive to the adopted inflow parameters of \hpri{} and \hsec{}. 

Since the inflow parameters of \hpri{} and \hsec{} have not been measured directly, we decided to check how the simulated signal reacts to variations in their speeds, inflow directions, and temperatures. 
We put a constraint on the variations of the adopted parameters such that they are varied in a correlated way, so that the weighted mean of the speeds and temperatures are maintained and equal to those obtained by \citet{lallement_etal:10b}. 
Also the directions of inflow of \hpri{} and \hsec{} are varied such that the weighted mean values remained unchanged, with an additional constraint that the varied directions slide along the B-V plane, obtained by \citet{kubiak_etal:16a} based on the inflow directions of the primary and secondary populations of ISN He. 
The rationale behind this approach is that the mean inflow direction is known relatively well from observations of the helioglow, both spectroscopic \citep{quemerais_etal:99, lallement_etal:05a, lallement_etal:10a, katushkina_etal:15a} and photometric \citep{koutroumpa_etal:17a, bzowski_etal:23b}. 
The mean direction for \hpri{} and \hsec{} adopted in our modeling agrees very well with these estimates.
The baseline parameters of \hpri{} and \hsec{} are listed in the first row of Table \ref{tab:shiftParam}. 
The varied parameters are listed in the subsequent rows in the table. 

\begin{deluxetable*}{|ll|llll|c|}[!ht]
\tablecaption{\label{tab:shiftParam}  Parameters of inflow of the primary and secondary populations of ISN H used in the simulations. }
\tablehead{nr &case & longitude [\degr] & latitude [\degr] &  speed [\kms] & temperature [K] &  step }
\startdata
0 & nominal \hpri    &  255.745 & 5.169 & 25.784 &  7443 & \\
 & nominal \hsec    &  251.570  & 11.95 & 18.744 & 16\,300 &  \\
\hline
1 & shiftOut \hpri   &  257.90   & 2.41  & 25.784 &  7443 & $\Delta \lambda_\text{BV,pri} = -3\fdg5$ \\
 & shiftOut \hsec   &  249.51  & 12.9  & 18.744 & 16\,300 & $\Delta \lambda_\text{BV,sec} = 2\fdg0 $ \\
\hline
2 & shiftIn \hpri   &  253.570  & 7.9   & 25.784 &  7443 &$\Delta \lambda_\text{BV,pri} = 3\fdg5$\\
 & shiftIn  \hsec   &  252.10   & 9.78  & 18.744 & 16300 & $\Delta \lambda_\text{BV,sec} = -2\fdg0 $ \\
\hline
3 & $\Delta$T \hpri   &  255.745 & 5.169 & 25.784 &  6300 &  $\Delta T_\text{pri}=-1143$ K \\
 & $\Delta$T  \hsec   &  251.57  & 11.95 & 18.744 & 17300 &   $\Delta T_\text{sec}= 1000$ K \\ 
\hline
4 & $\Delta V$ \hpri  &  255.745 & 5.169 & 27.784 &  7443 &  $\Delta V_\text{pri}= 2.0$ \kms   \\ 
 & $\Delta V$ \hsec   &  251.570  & 11.95 & 17.6   & 16300 &  $\Delta V_\text{sec} = -1.144$ \kms \\
\hline
5 & 		\hmix	 &     252.50   &   8.9 & 21.26   & 12860 &  
\enddata
\end{deluxetable*}

\subsubsection{Variation of the inflow directions}
\label{sec:varInflowDir}
\noindent
Case 1 (shiftOut) involves shifting the direction of inflow of \hsec{} by $\Delta \lambda_\text{BV,sec} = 2\degr${} away from the mean value in the B-V plane.
This results in a correlated change of the inflow direction of \hpri{} by $\Delta \lambda_\text{BV,pri} = -3\fdg5$ in the B-V plane. 
Converted back to the ecliptic coordinates, these new directions are listed in the first two columns of the second panel of Table \ref{tab:shiftParam}.
The speeds and temperatures remain unchanged.
The effect of this variation is a change in the angle between the inflow directions of \hpri{} and \hsec{} from 7\fdg88 to 13\fdg37. 

Case 2 in Table \ref{tab:shiftParam} (shiftIn) corresponds to a parameter variation in the opposite direction, so that the inflow directions in the B-V plane are closer to each other than in the nominal case. 
The longitude of the inflow direction of \hsec{} was reduced by 2\degr, which resulted in an increase of the B-V longitude of \hpri{} by 3\fdg5. 
The separation of the inflow directions became then equal to 2\fdg38.
Again, the speeds and temperatures remained unchanged. 

\begin{figure}
\centering
\includegraphics[height=0.41\textwidth]{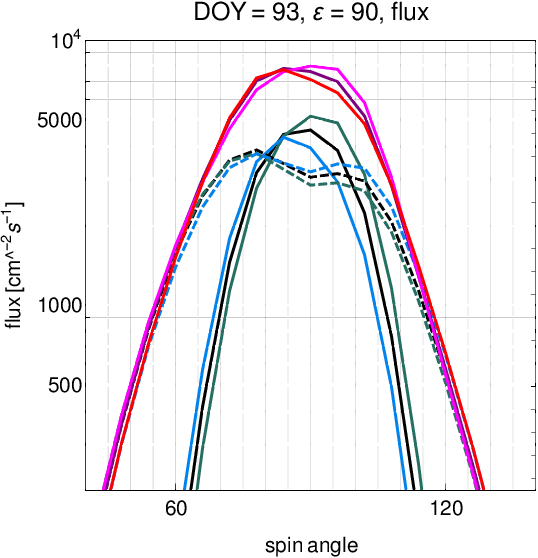}
\includegraphics[height=0.41\textwidth]{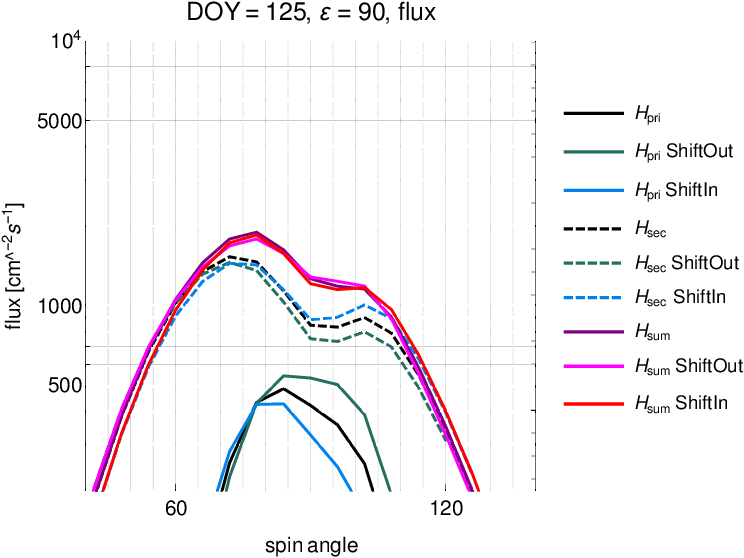}
\caption{
The figure presents simulated fluxes for \hpri, \hsec, and their sum (H$_\text{sum}$) for the nominal inflow parameters and those with the varied directions (shiftOut and shiftIn cases - Table \ref{tab:shiftParam}), as shown in the legend at the right side of the figure. 
The differences between the parameter sets are marked with different colors, and the line style (solid vs broken) differentiates between \hpri{} and \hsec. 
The purple, magenta and red colors mark the respective sums of \hpri{} and \hsec{} for the three sets of the directions of inflow. 
}
\label{fig:HshiftInOut}
\end{figure}

The results of these virtual experiments are presented in Figure \ref{fig:HshiftInOut} in conjunction with the electronic Figure 15 in the Appendix of  \citet{kubiak_etal:23a}, which shows that both of the presented (DOY, $\varepsilon$) combinations have no significant contributions from ISN He.
The simulated signals from individual populations visibly react to the changes of the inflow directions by a few degrees, but the sum of the two populations varied in the correlated way is much less sensitive.
Still, differences between the nominal case and the cases shiftOut, shiftIn are visible in the logarithmic scale. 
On the left panel \hpri{} and \hsec{} have comparable strengths throughout the center of the spin angle range, and at the wings \hsec{} dominates.

In a different example of viewing geometry, shown in the right panel of Figure \ref{fig:HshiftInOut}, the dominant population throughout the entire spin angle range is \hsec.
Also here differences between the fluxes for the three different inflow directions are well visible for individual populations \hpri, \hsec, but for the sums, the difference is much smaller. 
A reduction in the flux of one of the populations is compensated for by an increase in the other population as the inflow directions slide along the B-V plane such that the mean inflow direction remains unchanged.

This implies that the sensitivity of the signal due to \hsec{} for this viewing geometry to shifts in the inflow directions along the B-V plane is relatively small but might be detectable in actual observations if the statistics is good.
Note that the total flux of ISN H is expected to be one of the largest expected for the solar maximum conditions, and the energy conditions are fulfilled. 

The observability of this aspect is illustrated in Figure \ref{fig:doyEShift}.
The left panels in each row show with color dots the (DOY, $\varepsilon$) combinations for which the flux between the nominal case and the test case differences exceed 5\%, which makes them best suitable for analysis of this effect. 
The second panel represents the actual magnitudes of the flux ratios for those (DOY, $\varepsilon$) combinations that are marked with the color in the left panel.
The third panels represent examples of the fluxes. 

A conclusion from this part is that identifying the inflow directions of \hpri{} and \hsec{} will require a very careful analysis and observations for different combinations of (DOY, $\varepsilon$). 
The (DOY, $\varepsilon$) combinations well suitable for studies of the variations in the flow directions of \hpri, \hsec{} are presented in the first and second rows of Figure \ref{fig:doyEShift}.
The scenario for $\varepsilon$ settings suggested by \citet{kubiak_etal:23a} provides such opportunities.

\begin{figure}
\centering
\includegraphics[height=0.42\textwidth]{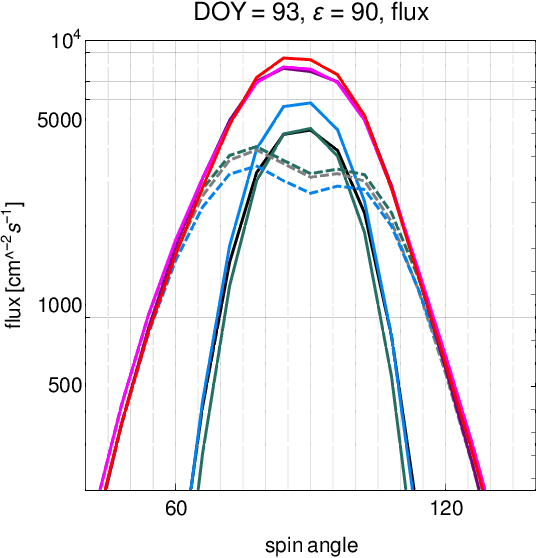}
\includegraphics[height=0.42\textwidth]{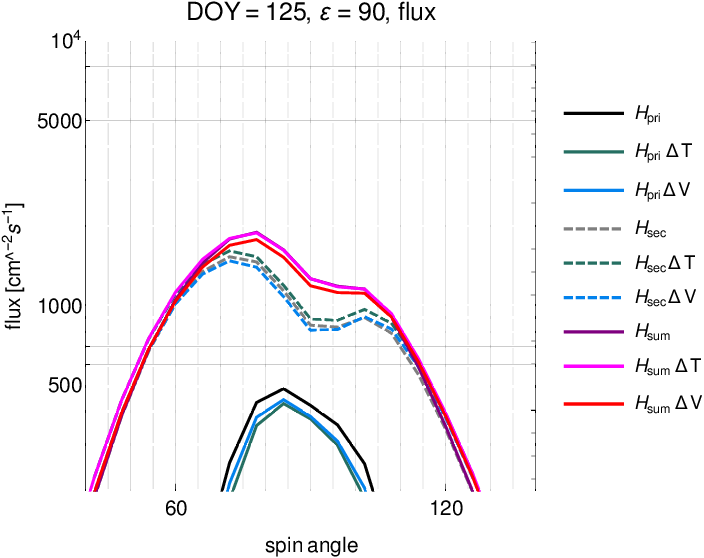}
\caption{
Effects of variation of the temperatures and speeds for the flux of \hpri, \hsec{} populations of ISN H and their sum, for the parameters listed in the third and fourth row in Table \ref{tab:shiftParam}. 
The combinations of (DOY, $\varepsilon$) are identical to those shown in Figure \ref{fig:HshiftInOut}.
The lines with $\Delta T$ in the legends corresponds to variations in the temperatures, and those with $\Delta V$ to those in speeds.
}
\label{fig:HshiftTV}
\end{figure}

\subsubsection{Variation of the temperatures}
\label{sec:temperaVar}
\noindent
Another test that we have made is for the sensitivity of the signal to temperature variations. 
We requested to maintain the speeds and inflow directions of \hpri, \hsec{} identical to those listed as the nominal case in Table \ref{tab:shiftParam}. 
Based on the temperatures of \hpri, \hsec{}, we calculated the mean thermal speed as a mean of the individual thermal speeds weighted by the population densities. 
Using this average thermal speed, we calculated the mean temperature, which was obtained 12\,680 K.

With this, we varied the temperatures of \hpri, \hsec.
For \hpri, we assumed a temperature of 6300 K, and maintaining the mean thermal spread of the two populations, we calculated the corresponding temperature of \hsec, equal to 17\,300 K.
This corresponds to the case 3, $\Delta T$ in Table \ref{tab:shiftParam}.

Results of this experiment are illustrated in Figure \ref{fig:HshiftTV} (see the cases with a denotation ``$\Delta T$'' in the legends). 
Similar as it was for the case of variations of the inflow directions, the individual populations are sensitive to the adopted temperature. 
When, however, the temperatures of \hpri{} and \hsec{} are varied so that the mean thermal spread is conserved, then H$_\text{sum}$ varies little. 
Still, they may be detectable in actual observations provided the statistics is sufficient.

The combinations of (DOY, $\varepsilon$) well suitable for studies of the temperature effects are presented in the third row of Figure \ref{fig:doyEShift}.

\subsubsection{Variation of the speeds}
\label{sec:speedVar}
\noindent
The last of the varied parameters were the inflow speeds of \hpri, \hsec.
We increased the inflow speed of \hpri{} by $\Delta V_\text{pri} = 2$~\kms{} and accordingly reduced that of \hsec{} such that the weighted mean speed remained unchanged (Case 4 in Table \ref{tab:shiftParam}).

\begin{figure}
\centering
\includegraphics[width=1\textwidth]{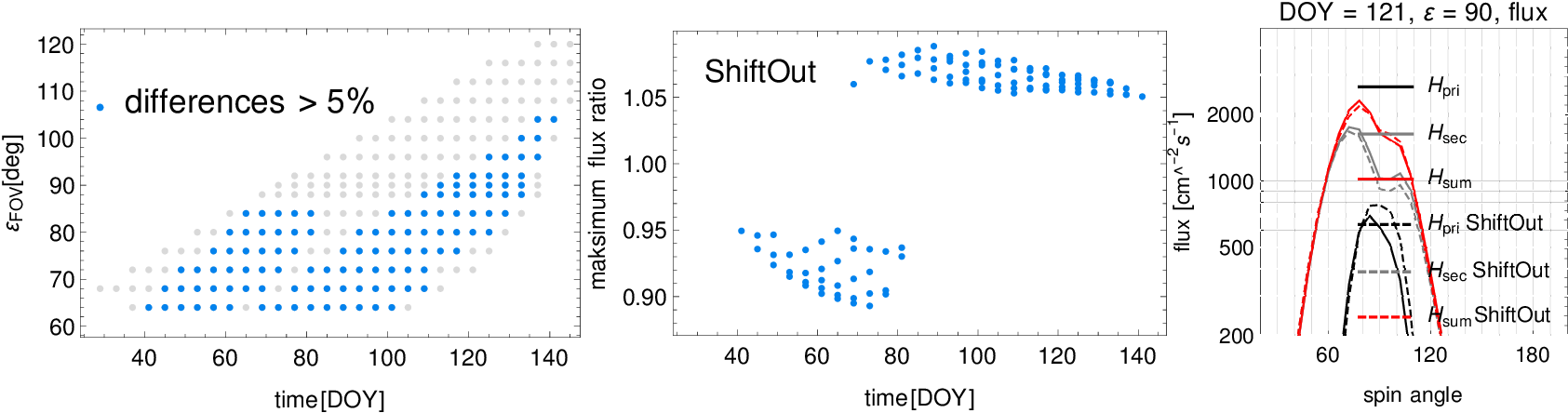}
\includegraphics[width=1\textwidth]{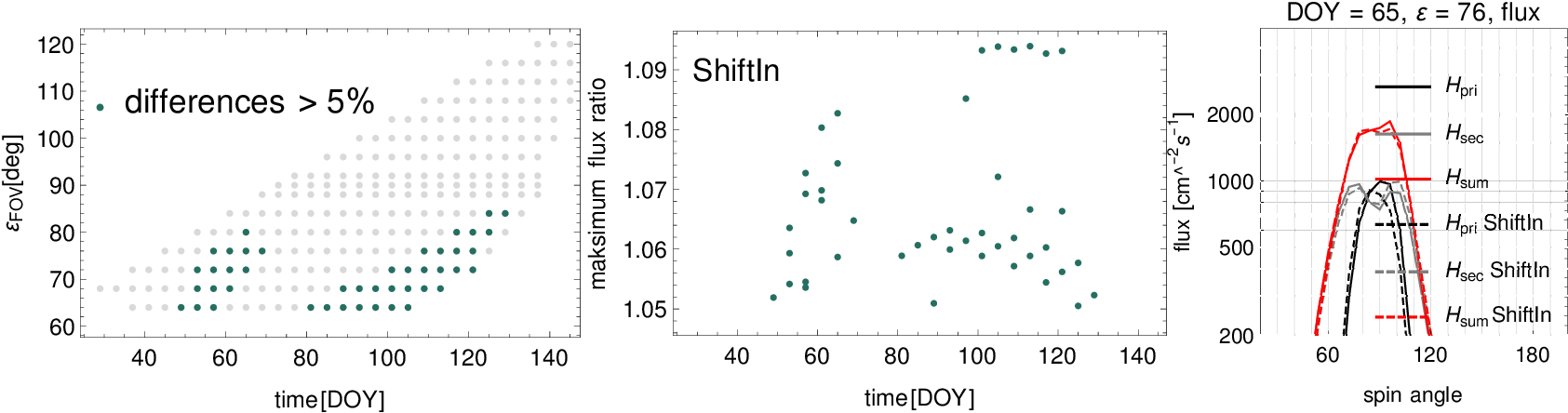}
\includegraphics[width=1\textwidth]{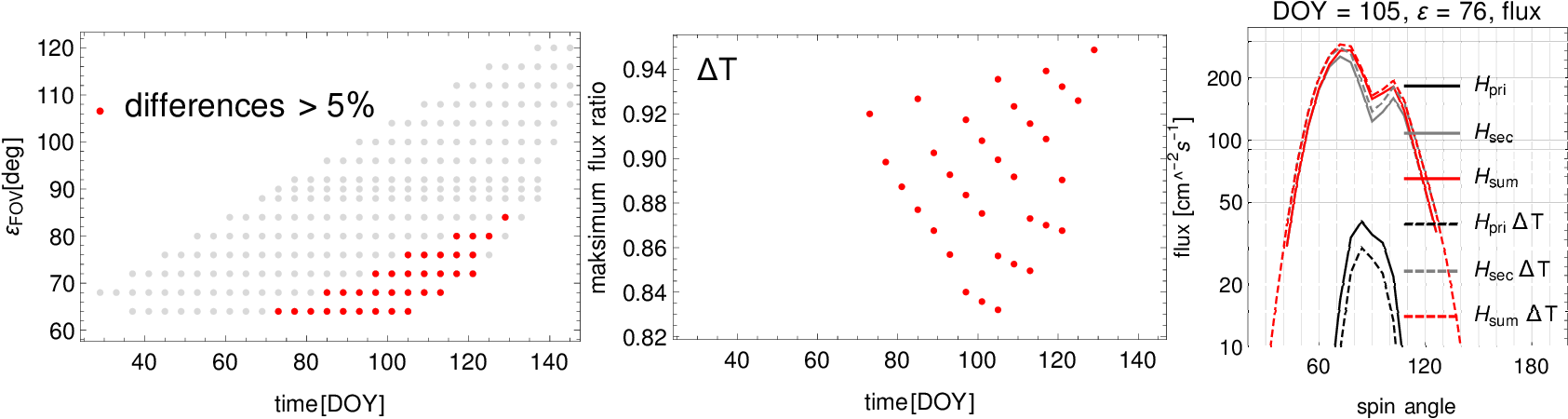}
\includegraphics[width=1\textwidth]{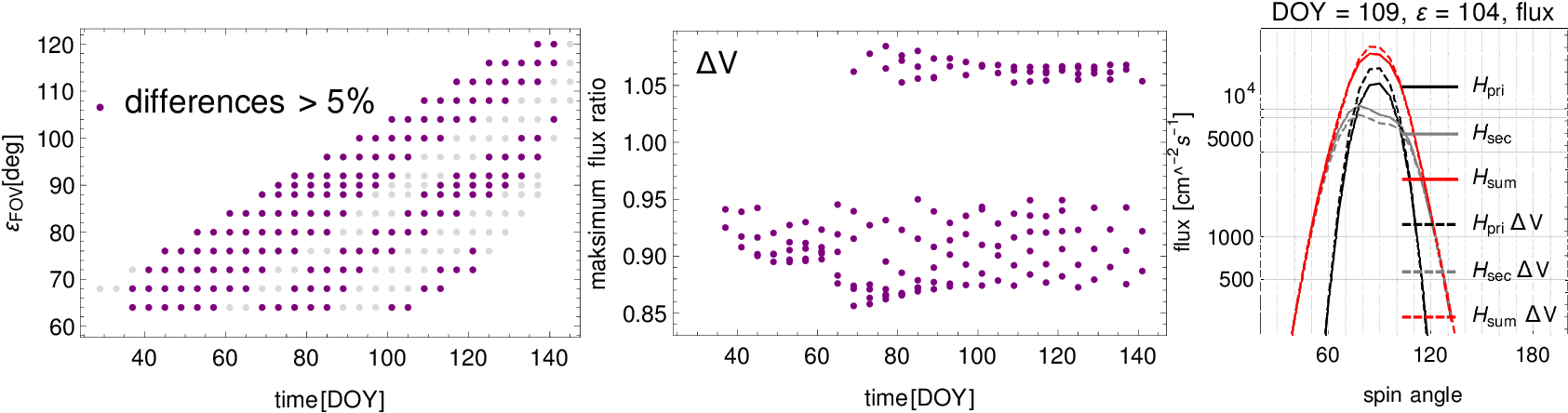}
\caption{
Locations in the (DOY, $\varepsilon$) space well suitable for studies of the flow directions (first and the second row), temperatures (third row), and speeds (fourth row) of individual populations of ISN H.
The leftmost column of panels presents the (DOY, $\varepsilon$) combinations for which the flux differences for a given parameter are expected to exceed 5\% either upwards or downwards (colored dots).
The gray dots represent the region in the (DOY, $\varepsilon$) space where the energy and flux thresholds are exceeded, and He$_\text{sum} < \hsum$. The colored dots are always on top of gray, because the condition He$_\text{sum} < \hsum$ for the colored dots is always fulfilled.
The center column presents the magnitude of the ratio of the flux maxima. 
Note that for a given DOY, the number of colored dots in the left panel is equal to that shown for this DOY in the center panel.
The rightmost column presents ISN population fluxes for example (DOY, $\varepsilon$) combinations for the selected effects.
The effects presented (from top to bottom) correspond to cases 1--4 in Table \ref{tab:shiftParam}. 
}
\label{fig:doyEShift}
\end{figure}

The effect of these changes on the simulated signal for selected (DOY, $\varepsilon$) combinations are shown in Figure \ref{fig:HshiftTV} (see the cases denoted with ``$\Delta V$'').
Again, changes in individual populations are well visible, but the total flux varies very little, because the changes in individual populations compensate each other almost entirely. 
Also note that the differences between the simulated flux for the case ''H$_\text{sum} \Delta T$'' and ''H$_\text{sum} \Delta V$'' vary little. 
Variations in speed of \hpri{} and \hsec{} gives larger differences in the flux than the variations in the temperature, discussed in Section \ref{sec:temperaVar}, which is especially well visile in the left panel of Figure \ref{fig:HshiftTV}.
The combinations of (DOY, $\varepsilon$) well suitable for studies of the velocity effects are presented in the fourth row of Figure \ref{fig:doyEShift}.

\subsubsection{Section summary}
\label{sec:hMixConclu}
\noindent
These numerical experiments demonstrate that the signal for ISN H is relatively little sensitive to considerable variations in the parameters of inflow of the primary and secondary populations as long as these variations maintain the mean values of the parameters. 
Nevertheless, a sufficient observation statistics is expected to allow for investigation of these effects. 
It seems interesting to study the parameters of inflow of \hsec{} for these (DOY, $\varepsilon$) combinations where the secondary population dominates. 
With such combinations identified for DOYs far apart during the year, it will hopefully be possible to apply the methodology proposed by \citet{bzowski_etal:22a} and determine the parameters of this population without parameter correlation. 
To that end, an appropriate variation of the elongation angle will be needed.
Then, since the mean parameters of inflow of ISN H are already known and hopefully will be established even more accurately based on IMAP measurements, it will also be possible to determine those of \hpri. 

\subsection{Sensitivity of the simulated flux of ISN H to one- vs two-Maxwellian approximation}
\label{sec:Hmix}
\noindent
In earlier studies, we verified that the one-Maxwellian model (1Max) reproduces very well the results obtained using the two-Maxwellian approximation (2Max) for the distribution of the heliospheric hydrogen glow observed from 1 au \citep{kubiak_etal:21b}. 
This implies that also the spatial distribution of the density of ISN H deep inside the heliosphere must be reproduced correctly.
And if the density is represented correctly, so must be the production of interstellar pickup ions inside the heliosphere. 

By contrast, in the case of He, \citet{bzowski_etal:12a} demonstrated that it is not possible to represent the signal due to ISN He measured by IBEX-Lo using the 1Max representation of the distribution function. 
Further analysis by \citet{kubiak_etal:14a} and \citet{kubiak_etal:16a} showed that the two-maxwellian approximation is sufficient to reproduce the observed signal. 
\citet{kubiak_etal:19a} showed that the distribution function of ISN He at 1 au simulated using the 2Max approximation differs very little from this function simulated using a more realistic model developed by \citet{bzowski_etal:17a}, where the production of the secondary population is simulated as a natural consequence of charge-exchange collisions in the outer heliosheath.

In this section, we check for the differences between 1Max and 2Max approaches to verify if IMAP-Lo will be able to resolve the two populations of ISN H.
The baseline model is the two-population approach. 
We compare the simulated signal with that for the 1Max model for the flow parameters obtained as weighted means of the parameters of \hpri{} and \hsec{}.
In this choice, we follow \citet{IKL:18b} and \citet{kubiak_etal:21b}.
The parameter values used in the 1Max approximation are listed in the last row of Table \ref{tab:shiftParam}.
We denote this simulation as the \hmix{} case.

\begin{figure}[!ht]
\centering
\includegraphics[width=0.65\textwidth]{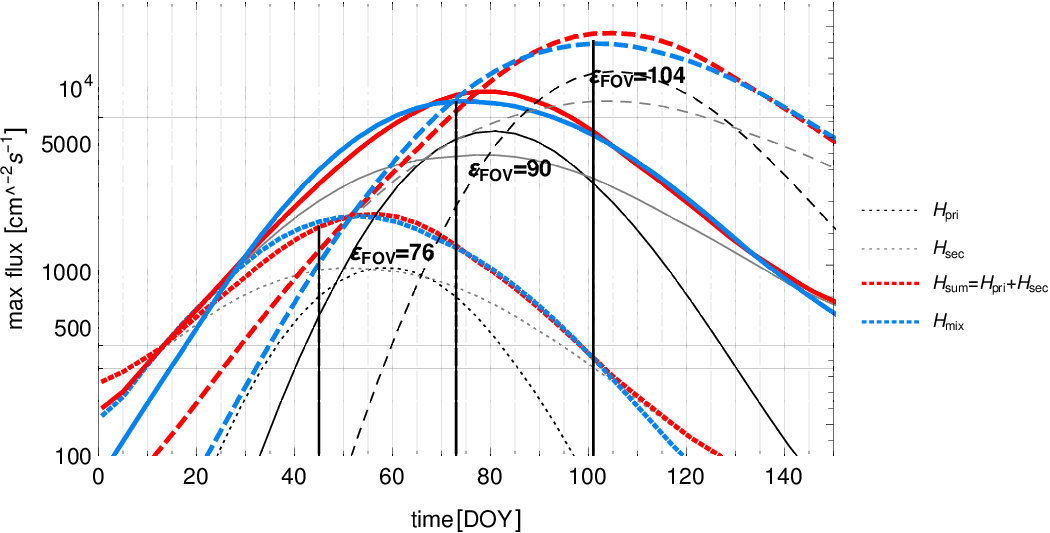}
\caption{
A comparison of the maxima of the flux of ISN H over spin angle for the three elongations proposed for the first year of IMAP-Lo operations: 76\degr, 90\degr, and 104\degr.
The black and gray colors represent \hpri{} and \hsec{} populations, respectively, and red marks their sum \hsum.
The 1Max approximation is represented by the blue color.
Different line styles represent the different $\varepsilon$ angles, as marked up in the plot. The black vertical lines refer to the days, when He$_\text{sum}$ drop below $\hsum$, and after that He does not cover the H measurements.
}
\label{fig:Hmix1}
\end{figure}

We performed simulations of the signal for the \hmix{} population on the same (DOY, $\varepsilon$) grid as the other simulations discussed in this paper.
We started this part of the analysis by comparing spin-angle maxima of the simulated fluxes as a function of DOY for the three lines in the (DOY, $\varepsilon$) space suggested by Ku 23 to be followed during the first year of the IMAP mission.
The results are presented in Figure \ref{fig:Hmix1}.
We show the population maxima vs DOY for \hpri{}{} and \hsec{}, but the essence of the analysis is in a comparison of the maxima for \hsum{} and \hmix, where the first represents maxima of the signal for $\hpri + \hsec$, and the other one for the 1Max population.
They are represented with the red and blue colors, respectively.
Clearly, the blue and red lines do not overlap, which suggests that in general, \hsum{} is visibly different from \hmix. 
In other words, we can investigate the two populations of ISN H based on the planned IMAP-Lo observations. 
Note that this conclusion holds even though, according to the discussion presented in Section \ref{sec:onlyH}, the signal due to ISN H will not be swamped in that due to ISN He only after $\sim$ DOY 45 for $\varepsilon = 76\degr$, and DOY 101 for $\varepsilon = 104\degr$. 
The largest differences are expected for $\varepsilon = 90\degr$ and 104\degr. 

\begin{figure}[!ht]
\centering

\includegraphics[width=0.32\textwidth]{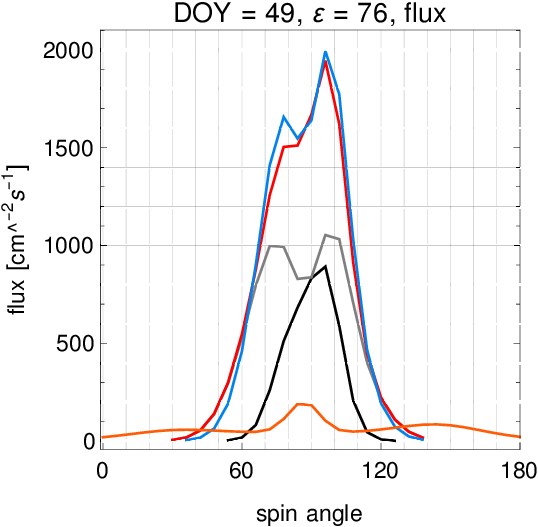}
\includegraphics[width=0.33\textwidth]{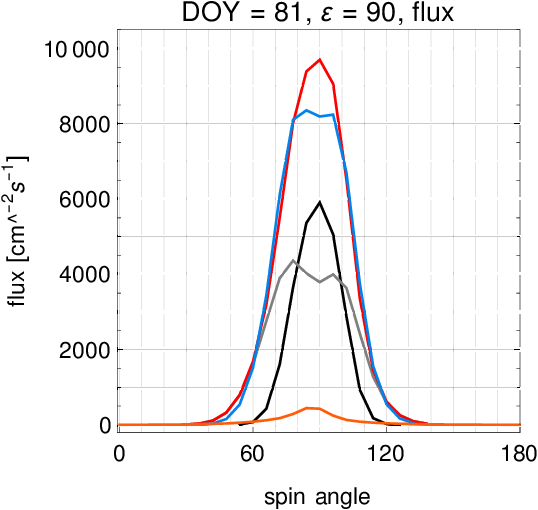}
\includegraphics[width=0.33\textwidth]{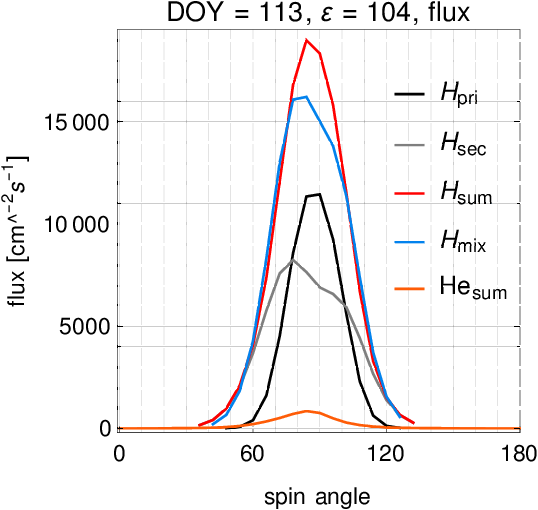}

\includegraphics[width=1\textwidth]{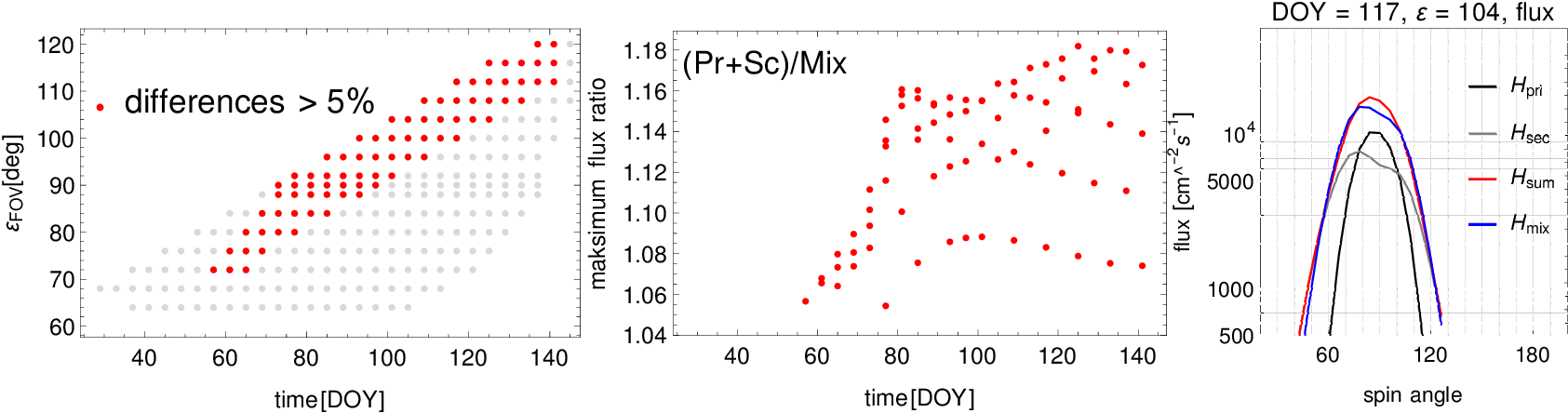}
\caption{
Upper row: example differences between the 1Max and 2Max cases for the distribution function of ISN H at the heliopause. 
The fluxes for the populations \hpri{} and \hsec{} and their sum \hsum{} are compared with the flux expected for the 1Max approximation, denoted as \hmix. 
The inflow parameters for \hpri, \hsec{} are given in the 0-th row of Table \ref{tab:shiftParam}, and for \hmix{} in the 5-th row in this table. 
Additionally, a sum of the fluxes for the primary and secondary populations of ISN He is presented, which shows that ISN He is not expected to hamper the analysis because the expected flux of ISN He is negligible for these combinations of (DOY, $\varepsilon$).  
Note the different shapes of flux vs spin angle for the 1Max and 2Max cases.
The regions in the (DOY, $\varepsilon$) space where the differences between the fluxes simulated for \hmix{} and \hsum{} exceed 5\% are shown in the lower left panel, the corresponding magnitudes of the  differences in the lower center panel, and a comparison of the fluxes vs spin angle for example (DOY, $\varepsilon$) pair (right column) in the format identical with that used in the preceding sections in the lower right panel.
}
\label{fig:Hmix2}
\end{figure}

An illustration of various aspects of differentiation between the 1Max and 2Max cases is presented in Figure \ref{fig:Hmix2}. 
In the first panel of the upper row, the difference between the 1Max and 2Max cases is very small.
However, for the examples shown in the second and third panel of the first row, the differences between the 1Max and 2Max cases are clearly visible, especially near the peak of the flux, where the statistics is expected to be the best.  
Note the different proportions between the peak magnitudes in the 1Max and 2Max cases between the three panels.
Also note that the figures are in linear scale, unlike most of the flux figures in the paper, and that the range of the vertical scales differs between the panels. 
All this implies that it is possible to differentiate between various approximations to the distribution function ISN H within the suggested scenario. 

The regions in the (DOY, $\varepsilon$) space where these differences are expected to be the largest are presented in the lower row of panels in Figure \ref{fig:Hmix2}.
The peaks for the \hmix{} case are consistently lower than those for \hsum{} = \hpri{} + \hsec.
They are best visible for elongations between 72\degr{} and 120\degr, for DOYs between $\sim 60$ and 140.  
The magnitudes of the differences typically exceed 10\%, as shown in the center panel of the lower row. 
The extreme right panel in the lower row presents another example of differences between the cases discussed, this time in the format identical to that used in most of the flux figures. 
This is to provide correspondence with the discussion presented in the other sections. 

\subsection{Sensitivity of the H signal to the dependence of radiation pressure on radial velocity of atoms}
\label{sec:VrH}
\noindent
The strength of the radiation pressure force acting on individual H atoms depends on the magnitude of the solar spectral flux at the wavelength corresponding to the Doppler shift of the Lyman-$\alpha$ wavelength due to the non-zero radial velocity of the atom relative to the Sun, as shown in Figure \ref{fig:HDRadPress}. 
\citet{tarnopolski_bzowski:09} pointed out the significance of this effect for the density distribution of ISN H in the inner heliosphere. 
\citet{IKL:18b} and \citet{IKL:20a} analyzed differences in the simulated signal for the observation conditions of IBEX when using different models of the solar spectral flux in the Lyman-$\alpha$ line. 

\begin{figure}[!ht]
\centering
\includegraphics[height=0.27\textwidth]{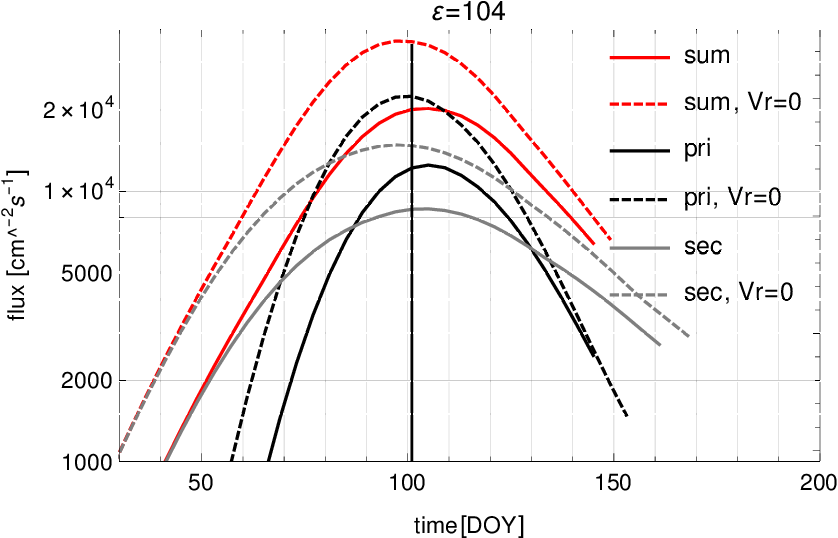}
\includegraphics[height=0.27\textwidth]{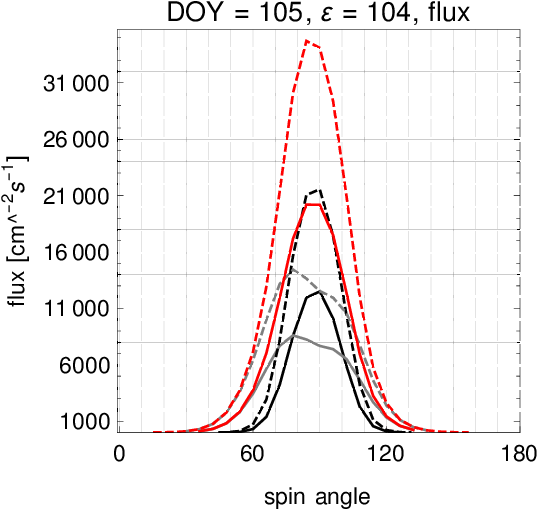}
\includegraphics[height=0.27\textwidth]{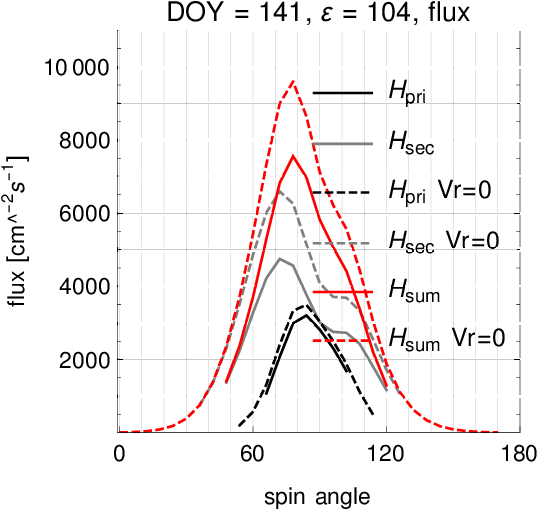}

\includegraphics[width=1\textwidth]{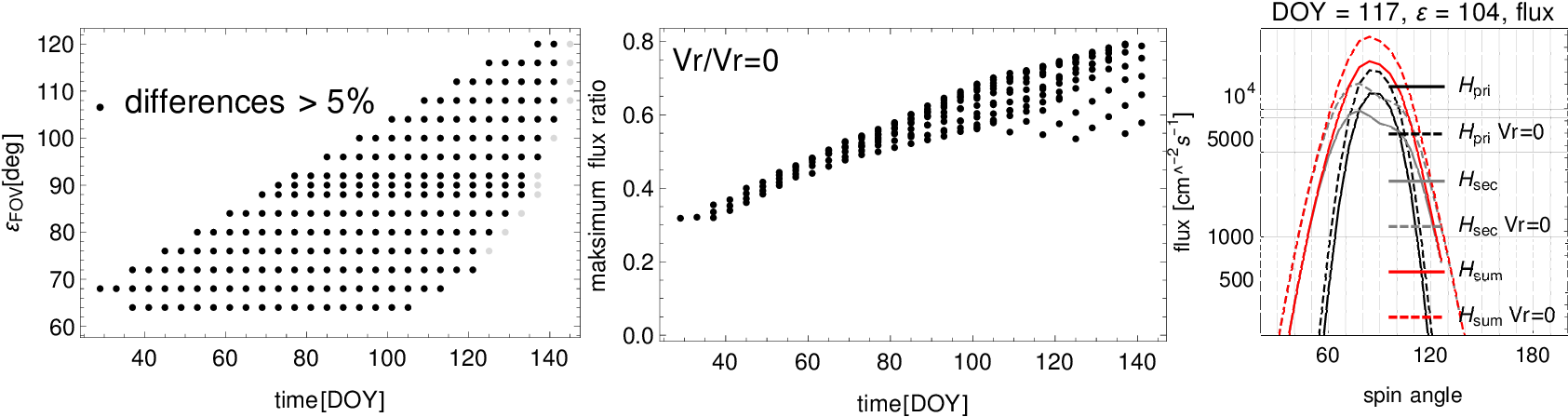}
\caption{
Upper left: magnitudes of the spin angle-maximum flux of \hpri, \hsec, and \hsum{} as a function of DOY in 2015, $\varepsilon = 104\degr$ for a realistic model of radiation pressure, with the $\mu$ factor being a function of both time and radial speed of the atom (solid lines) and for a model with $\mu$ independent of $v_r$ ($\mu(v_r) \equiv \mu(0)$; broken lines). 
Upper center and right: Flux vs spin angle for two example (DOY, $\varepsilon$) pairs for \hpri{}, \hsec{}, and \hsum{} for radiation pressure dependent (solid) and independent of $v_r$ (broken lines) in the linear scale. 
The lower row illustrates the locations in the (DOY, $\varepsilon$) space where the differences between the simulations with and without dependence of radiation pressure on $v_r$ are larger than 5\% (left), the corresponding magnitudes of the differences (center), and fluxes vs spin angle for another example of (DOY, $\varepsilon$) pair in the log scale (right). 
}
\label{fig:VrH}
\end{figure}

\citet{rahmanifard_etal:19a} used a radiation pressure-independent model of the IBEX-Lo signal due to ISN H and pointed out that most likely, the effective radiation pressure acting on the ISN H atoms detected by IBEX-Lo is different to that obtained by \citet{IKL:18a} based on direct observations of the solar Lyman-$\alpha$ line by \citet{lemaire_etal:15a}. 
\citet{katushkina_etal:21b} used a simulation model of the IBEX-Lo signal due to ISN H with the radial speed dependence of radiation pressure included based on \citet{kretzschmar_etal:18a} and suggested that the global simulation model of the heliosphere they had used to simulate the signal was unable to reconcile the simulations with the measurements unless the actual profile of the solar Lyman-$\alpha$ line was steeper than that suggested by \citet{IKL:18a}.

Here, we study the sensitivity of the expected signal observed by IMAP-Lo to the dependence of radiation pressure on radial velocity of individual atoms. 
We compare the fluxes simulated assuming the radiation pressure obtained from the model by \citet{IKL:20a}, i.e., with radiation pressure $\mu$ factor dependent both on time and radial velocity, with that with the time-dependent magnitude of radiation pressure obtained for $v_r = 0$.

We found that even though the magnitudes of the $\mu$ factor due to $v_r$-dependence do not differ strongly from those with the $v_r$ dependence neglected, their effect on the simulated flux of ISN H is large.
Adopting the radiation pressure characteristic for $v_r = 0$ and allowing no dependence on $v_r$, we obtained a flux magnitude almost twice larger than that obtained for a more realistic case with the dependence of $\mu$ on $v_r$ included, for all (DOY, $\varepsilon$) pairs in the observation scenario suggested by \citet{kubiak_etal:23a}, as shown in Figure~\ref{fig:VrH} (top left and bottom center panels). 
Clearly, not only the magnitude of the flux is different, 
but also the day-to-day gradients of \hsum{} for these two cases are different.
Moreover, the times for the yearly maximum of the flux differ between these two cases, as well for \hsum{} as for the individual populations. The difference is visible at a level larger than 20\% for the entire (DOY, $\varepsilon$) set available for observations of ISN H (see the bottom left panel of Figure~\ref{fig:VrH}).

\section{Deuterium}
\label{sec:D}
\noindent
The abundance of ISN D at 1 au relatively to both ISN H and He is expected to be different than that in the unperturbed LISM \citep{tarnopolski_bzowski:08a}. 
The abundance relative to He is expected to be reduced.
The reason is on the one hand, the ionization losses for D are larger due to the larger ionization rate per atom, and on the other hand, the amplification of the flux at 1 au due to acceleration inside the heliopause is lower for D than for He.
This is because for He, radiation pressure is negligible \citep{IKL:23a}, and for D it is approximately equal to half of the solar gravitation force \citep{tarnopolski_bzowski:08a, IKL:18a}.
As a result, the losses of ISN D inside the heliosphere are larger than those of ISN He, and the D/He abundance at 1 au is reduced.

The abundance D/H at 1 au is expected to be increased.
This is because the gravitational amplification of the flux is larger for D than for H since radiation pressure for H is approximately two-fold greater than for D, and the solar gravity force is fully compensated. 
And while the ionization rates per atom are identical, the exposure time for D is shorter than that for H because the travel time is shorter, and consequently the ionization losses for D are smaller than those for H. 

This discussion is provided to remind the reader that the interpretation of the measurements must be performed taking into account the modification of the unperturbed abundances in the LISM inside the heliosphere. 
Another caveat is the fact that only the neutral portion of the local interstellar matter is able to penetrate the heliopause, so completing the measurement of the abundance in the LISM must take into account the ionization state of the matter, which is different for different species.
This can be addressed based on measurements of ISN He supplemented with appropriate modeling, as it was done, e.g., by \citet{bzowski_etal:19a}. 

\subsection{Detection method}
\label{sec:detectionMethod}
\noindent
ISN D is registered in the form of D$^-$ ions that are created by extraction of an electron by the impacting D atom from IMAP-Lo`s conversion surface.
Analysis of ISN D is challenging for two main reasons.
The first one is the very low flux impacting the instrument \citep{tarnopolski_bzowski:08a, kubiak_etal:13a}. 
This is because the abundance of ISN D in the LISM is low, about 15 ppm.
The other reason is the issue of separation of interstellar D$^-$ ions registered by the instrument from the terrestrial ones. 

As pointed out by \citet{rodriguez_etal:13a}, there are two sources of D$^-$ ions registered by IBEX-Lo: sputtering of D$^-$ by ISN He atoms from the terrestrial water layer covering the conversion surface within the instrument, and the genuine ISN D atoms converted into D$^-$ ions by electron capture at the conversion surface.
While ISN He sputters H$^-$ and D$^-$ ions from the terrestrial water layer, ISN D is not expected to sputter terrestrial-water H$^-$ and D$^-$ ions in significant amounts.

During the times when the observed signal is mostly due to the sputtering of H$^-$ by ISN He atoms, the ratio of D$^-$/H$^-$ counts is identical to the abundance of D in terrestrial water
. 
But when the instrument is impacted by ISN D atoms, then a certain amount of D$^-$ ions is created due to electron capture by these atoms appears in the signal. 
When the number of these ions becomes comparable to that of the sputtered D$^-$ ions, the observed ratio of D$^-$/H$^-$ changes. 
A positive detection of ISN D is achieved when the D$^-$/H$^-$ abundance differs from the terrestrial value statistically significantly.
This happens when on the one hand, there is a sufficiently large flux of ISN D at the instrument, and on the other hand, the flux of ISN He is strongly reduced.

Analysis of time-of-flight data from IBEX-Lo revealed three atoms of ISN D, accumulated over three observation seasons during the minimum of solar activity \citep{rodriguez_etal:13a, rodriguez_etal:14a}.
A much larger number of D$^-$ ions were found in the data, but they were attributed to the terrestrial water layer covering the conversion surface.
We investigate if the greatly enhanced observation capabilities of IMAP-Lo owing to the capability of adjustment of the elongation angle can enable observation of ISN D also during the solar maximum conditions. 

\subsection{Flux of ISN D}
\label{sec:Dflux}
\begin{figure}
\centering
\includegraphics[width=0.49\textwidth]{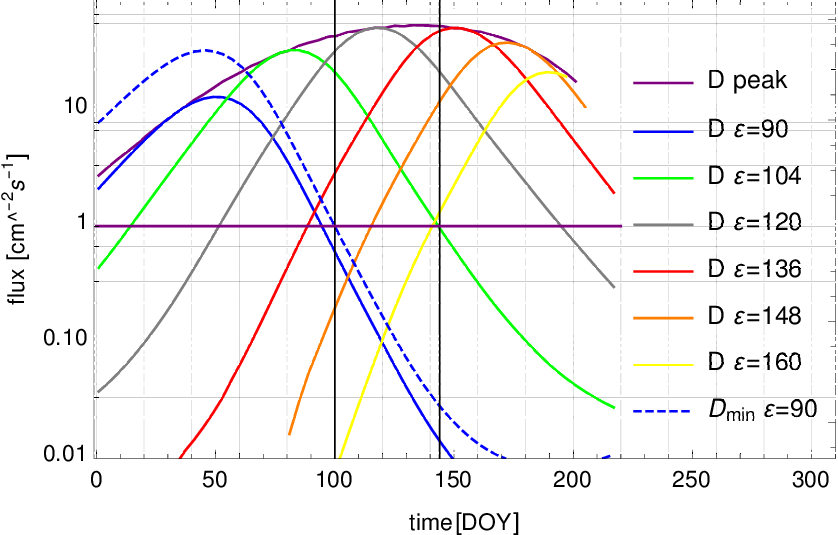}
\caption{
\noindent
Maxima of the flux of ISN D over spin angle, simulated for solar maximum conditions (year 2015) for the elongation angles $\varepsilon$ cover those used in the proposed scenario of IMAP-Lo observations \citep{kubiak_etal:23a}, color-coded (see the legend). The broken line marks the corresponding flux maxima for the IBEX-like viewing geometry ($\varepsilon = 90\degr$), solar minimum conditions (2009). The solid purple line marked ``D peak'' represents absolute daily maxima of the D flux in the entire $\varepsilon$ space.
For the horizontal and vertical bars, see the text.  }
\label{fig:Dflux}
\end{figure} 
\noindent
Similarly as on IBEX-Lo, we expect to be able to identify ISN D during the times when the beam of ISN He is absent in the field of view. 
Figure~\ref{fig:Dflux}, inspired by Figure 12 in \citet{sokol_etal:19c}, presents a comparison of the maxima of the fluxes of ISN D over spin angle for various elongation angles during a calendar year for an epoch of high solar activity (specifically, for the year 2015).  
The elongation angles shown are chosen to cover the $\varepsilon$ angles in the observation scenario suggested by \citet{kubiak_etal:23a}.
In addition, we show the magnitudes of the absolute maxima of the expected flux of ISN D as a function of the DOY.
For comparison, we show the peak fluxes of ISN D for the IBEX-Lo viewing geometry for 2009, i.e., for solar minimum conditions.

Clearly, owing to the ability of adjusting the elongation angle, we expect the flux of ISN D observed by IMAP-Lo during solar maximum to be of a similar magnitude to that observed by IBEX-Lo during solar minimum conditions. 
The effect of the increased solar activity for ISN D can be evaluated by comparison of the blue solid and broken lines in Figure \ref{fig:Dflux}.
Since we suggest performing the observations for a number of fixed elongations, it is interesting to identify those for which the expected flux of ISN D is at least as large as that for IBEX-Lo. 
These elongations are equal to 104\degr, 120\degr, and 136\degr.
However, during the times of the year when the expected flux for these elongations is the largest, $10-15$ \cmsq \persec{} sr$^{-1}$ , the flux of ISN He is on the order of $10^6$ \cmsq \persec{} sr$^{-1}$, so due to the reasons explained earlier, the flux of ISN D registered by IMAP-Lo will be swamped in terrestrial D$^-$ ions sputtered from the conversion surface by ISN He atoms.

\subsection{Counting rate of ISN D}
\label{sec:crD}
\noindent
To find the best strategy to look for ISN D within the suggested observation scenario \citep{kubiak_etal:23a}, we calculated the expected counting rate of D based on the simulated fluxes and their conversion to the counting rate identically as it was done by \citet{sokol_etal:19c}.
To that end, we used Equations 1 and 2 from the latter paper for the ISN and sputtered D, respectively.
The counting rate \cD{} of D$^-$ ions due to ISN D was calculated as a product of the total flux (i.e., as a sum of the fluxes of the primary and secondary populations) and the geometric factor, adopted as identical to that of IBEX-Lo , to simplify the comparisons. 
Note that the counting rate thus obtained is a conservative estimate because the field of view of IMAP-Lo will be larger than that of IBEX-Lo, so the geometric factor is expected to be greater by a factor $\sim 3.6$.
For the counting rate \cDterr{} of D$^-$ ions due to sputtering from terrestrial water layer covering the conversion surface by the impacting ISN He atoms, we take a product of the simulated flux of ISN He, the abundance of D in terrestrial ocean water \citep{boehlke_etal:05a}, and the geometric factor of IBEX-Lo for He. 
Details and the magnitudes of the coefficients are presented by \citet{sokol_etal:19c}.

\begin{figure}[!ht]
\centering
\includegraphics[width=0.49\textwidth]{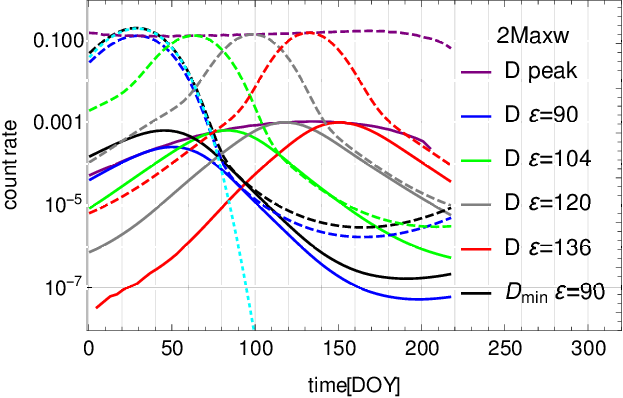}
\includegraphics[width=0.49\textwidth]{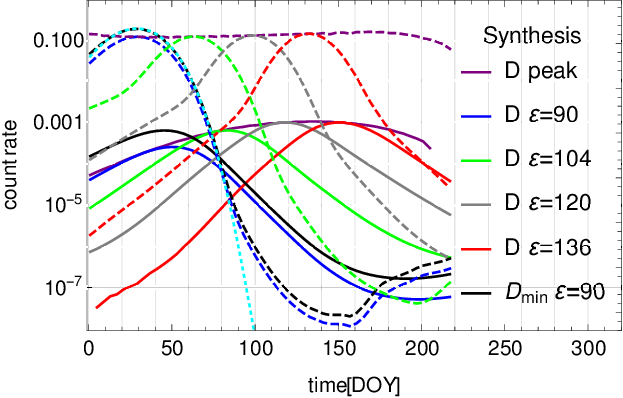}
\caption{
Simulated counting rates \cD{} due to ISN D (solid lines) and to the terrestrial D sputtered by ISN He \cDterr{} (broken lines) for selected elongations as a function of DOY.
The purple lines (D peak in the legend) correspond to the absolute maxima of the counting rates in the $\varepsilon$ space for individual DOYs for the ISN (solid) and sputtered terrestrial ions (broken line). 
The left panel presents the terrestrial sputtered D signal for simulation of the parent ISN He using the two-Maxwellian model, and the right panel that for the parent ISN He simulated using the synthesis method (see text). 
Note that unlike in \citet{kubiak_etal:13a} and \citet{rodriguez_etal:13a}, the sputtering flux of ISN He includes both the primary and secondary components of ISN He, which extends the time during the year when the sputtering of terrestrial D$^-$ ions is important. 
The case used by \citet{rodriguez_etal:13a} with only the primary ISN He used to calculate \cDterr{} is shown by the cyan line.
In the two panels, the solid lines are identical. Different are the broken lines, representing the counting rates of D$^-$ ions sputtered by ISN He. 
While the peaks of the broken lines are almost identical, the largest differences are at the far right slopes of the broken lines. 
}
\label{fig:mainD1}
\end{figure}

\begin{figure}[!ht]
\centering
\includegraphics[width=0.49\textwidth]{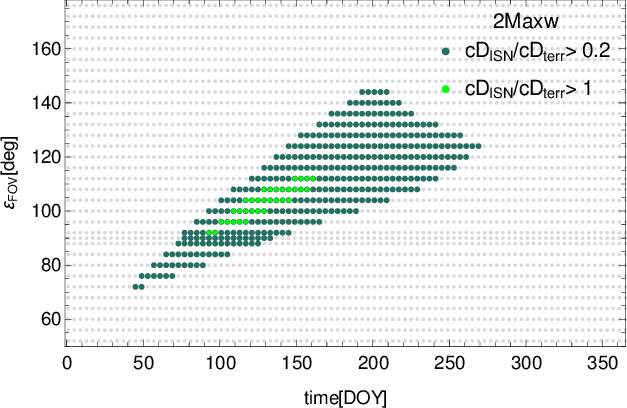}
\includegraphics[width=0.49\textwidth]{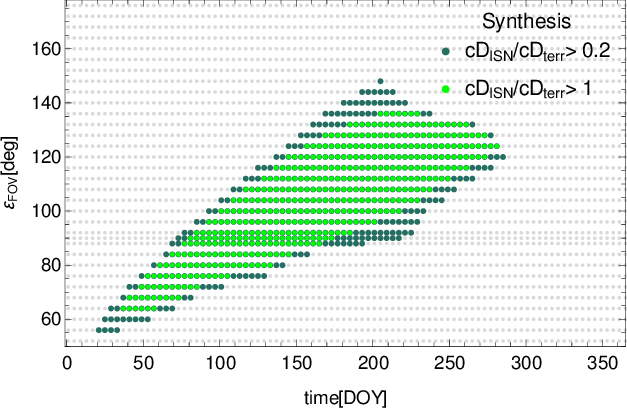}
\caption{
Regions in the (DOY, $\varepsilon$) space where the counting rate \cD{} of IMAP-Lo due to ISN D is larger than 20\% of that due to the terrestrial D sputtered by ISN He (\cDterr, dark-green dots) and those where $\cD > \cDterr$ (light-green dots).
The small gray dots represent the grid in the (DOY, $\varepsilon$) space of available simulations.
The left panel represents the case for the two-Maxwellian model of the parent ISN He, and the right panel that of the synthesis method.
}
\label{fig:mainD}
\end{figure}

To estimate the sputtering flux of ISN He, we used two alternative approaches.
In the first one, we assumed that in the source region for ISN He, there are two populations of ISN He, the primary and secondary, given by the Maxwell-Boltzmann distribution functions with different parameters \citep[see][]{kubiak_etal:23a}.
This implies that the primary and secondary populations are independent of each other and do not interact in any way.
This approached was previously used by \citet{kubiak_etal:14a} and \citet{kubiak_etal:16a} and demonstrated to reproduce IBEX-Lo observations very well for a subset of data used by these authors in the data analysis. 
However, simulations of the flux of ISN He for the IBEX-Lo orbits for which ISN D is expected have never been performed before.
The alternative model for ISN He is the so-called synthesis method, developed by \citet{bzowski_etal:17a} and \citet{bzowski_etal:19a}, and subsequently used by \citet{kubiak_etal:19a} and \citet{swaczyna_etal:23c}.
In this approach, it is assumed that the distribution function of ISN He is a Maxwell-Boltzmann function far ahead to the heliosphere (at $\sim 1000$ au). 
Charge-exchange reactions in the outer heliosheath, responsible for the creation of the secondary population, are taken into account using the characteristics method of solution of the Boltzmann equation and appropriate production and loss equations. The spatial distribution, temperature, and bulk velocities of the parent interstellar He$^+$ ions in the outer heliosheath, needed in this approach, are taken from a global model of the heliosphere.
Here, we used the model of the heliosphere by \citet{heerikhuisen_pogorelov:10a}, identical with that used by \citet{bzowski_etal:19a}.

We found that the predictions of the two models for the flux of ISN He at IMAP for the observation geometries suitable for detection of ISN D differ very strongly, as illustrated in Figure \ref{fig:mainD1}. 
In this figure, we present with solid lines the maxima of the simulated counting rates of D$^-$ due to ISN D for the $\varepsilon$ angles equal to 90\degr, 104\degr, 120\degr, and 136\degr,
and in dashed lines the counting rates due to sputtering of D$^-$ by ISN He for the two models: two Maxwellian in the left panel, and the synthesis method in the right panel. While the peaks of the sputtered counting rates are almost identical for the two methods of simulations of He, the differences in the right-hand slopes, where an actual detection of ISN D is expected, are very large.
It is impossible to predict which of the two models is closer to the reality without making actual observations. 
These differences are important because they result in very different viewing conditions for ISN D.
Note, however, that the secondary population of ISN He was not taken into account by \citet{rodriguez_etal:13a}. 
These authors only considered the primary population, represented in Figure \ref{fig:mainD1} by the broken cyan line.
This needs to be compared with the black dashed lines, representing the counting rate simulated for the solar minimum conditions with the secondary population taken into account.
Clearly, the secondary population of ISN He will add an extra background of the sputtered terrestrial D in the time interval DOY 80--100. 
Due to the large uncertainty of the secondary population of ISN He for this viewing geometry, it is challenging to estimate the actual magnitude of this background.  

The prospective time intervals for detection of ISN D for a given elongation are those when the solid and broken lines are of comparable magnitudes  (see Figure \ref{fig:mainD1}). 
Almost everywhere, $\cDterr \gg \cD$, but for $\varepsilon = 90\degr$ and 104\degr{} there are time intervals when \cD{} and \cDterr{} become comparable. 
They are near DOY 130 for $\varepsilon = 104\degr$ (green line) and DOY 100 for $\varepsilon = 90\degr$ (blue line). 

The regions in the (DOY, $\varepsilon$) space best suitable for detection of ISN D are searched for by analyzing the ratios of the counting rates due to ISN D to those due to the sputtering of terrestrial D by ISN He atoms. 
Figure \ref{fig:mainD} presents maps of the regions in the (DOY, $\varepsilon$) space prospective for detection of ISN D. 
The figure was inspired by the lowermost panel of Figure 7 in \citet{sokol_etal:19c}. 
The counting rates were calculated as 
\begin{equation}
\cD = \sum \limits_{\psi_i = 55\degr}^{\psi_i = 115\degr} c_{\text{D, ISN}i} \Delta \psi,
\label{eq:cDdef}
\end{equation}
where \cD$_i$ is the counting rate for the spin angle bin $\psi_i$, and $\Delta \psi = 6\degr$ is the width of the spin angle bin. 
The other rates used in the comparisons are calculated similarly.
To qualify for Figure \ref{fig:mainD}, we used two alternative criteria for the ratios \cD{}/\cDterr{}: we requested this ratio to be $>0.2$ (dark green) or $>1$ (light green).
Clearly, the region potentially suitable for detection of ISN D is quite large, even though the simulations represent the time of high solar activity. 
This is true even for the two-Maxwellian case illustrated in the left panel of Figure \ref{fig:mainD}. For the synthesis-method case, where the region of high counting rate \cDterr is much smaller, the viewing conditions for ISN D are even better. 
We consider the two-Maxwellian case as the lower limit and show that even then, there are good reasons to believe that IMAP-Lo will be able to detect ISN D even in observations performed during high solar activity.

In the following, we only discuss the less-favorable case of the two-Maxwellian approximation and demonstrate that even for this case, IMAP has good prospects for observing ISN D.

\subsection{Ratio of counting rates}
\label{sec:cRatio}
\begin{figure}
\centering
\includegraphics[width=0.49\textwidth]{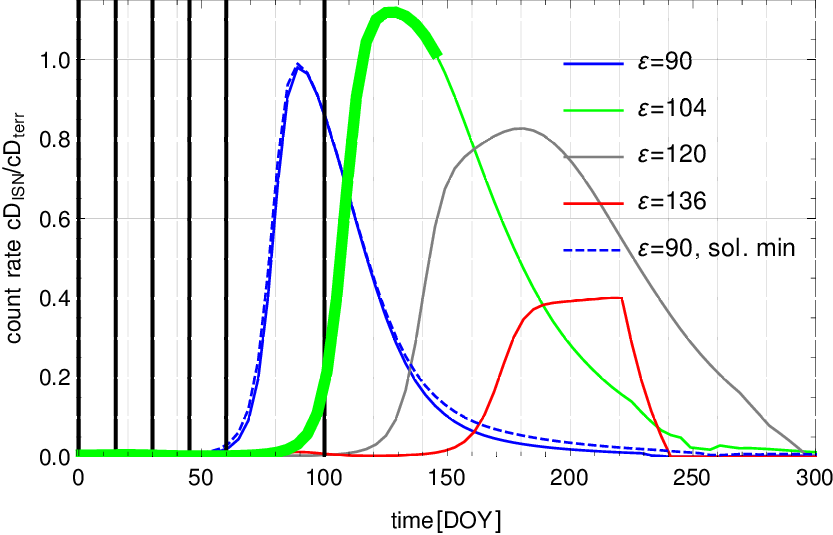}
\caption{
Evolution of the ratio \cD/\cDterr{} for several elongations during a portion of the year. 
Solid lines represent the conditions expected for IMAP (solar maximum), broken line those for the solar minimum conditions. 
The blue broken line represents the results of simulations for the viewing geometry of IBEX-Lo. 
The black vertical bars correspond to the five intervals used in the analysis of IBEX-Lo observations by \citet{rodriguez_etal:13a}. 
For the thick green line -- see text. The red line has a different shape due to the energy threshold.
The sputtered counting rate used was calculated based on the two-Maxwellian approximation. }
\label{fig:spinD}
\end{figure}

\begin{figure}
\centering
\includegraphics[width=0.48\textwidth]{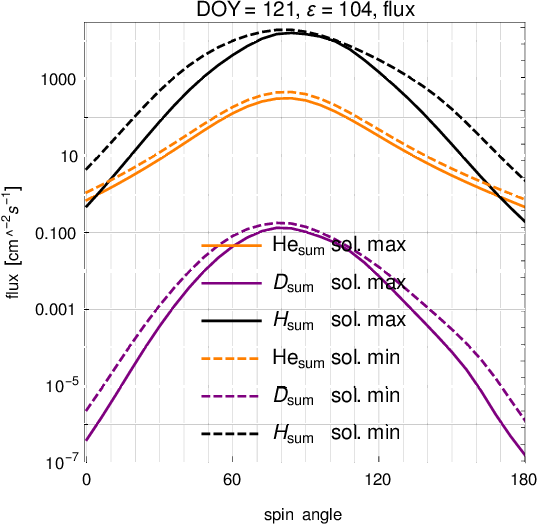}
\includegraphics[width=0.491\textwidth]{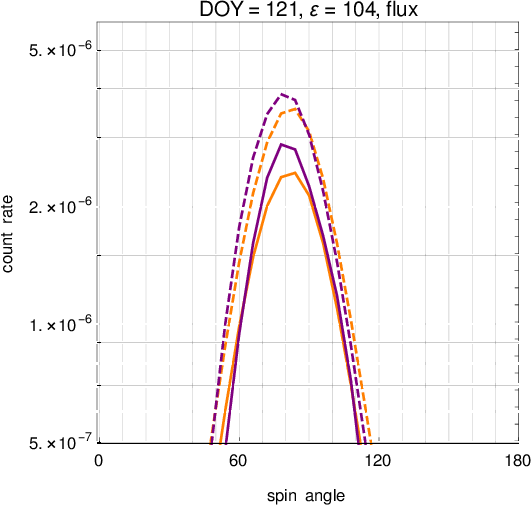}
\caption{
Fluxes (left) and counting rates (right) due to ISN D (purple) and the terrestrial D$^-$ sputtered by ISN He (orange) for the solar maximum (solid) and solar minimum conditions (broken lines), shown as a function of spin angle, for DOY 121, $\varepsilon = 104\degr$. 
In the left panel, also the flux of ISN H for the solar maximum and minimum conditions are shown. 
}
\label{fig:spinD1}
\end{figure}

\noindent
The results of the analysis by \citet{rodriguez_etal:13a} suggest that the most important quantity for successful detection of ISN D is the ratio of the counting rates $\cD/\cDterr$, and not the absolute magnitudes of the fluxes.
We checked the magnitude of this ratio for various elongations planned in the scenario proposed by \citet{kubiak_etal:23a}. 
They are presented in Figure \ref{fig:spinD}. 
The DOY range 60---100 corresponds to that for which \citet{rodriguez_etal:13a} identified three D atoms in the data collected during three years of observations of IBEX-Lo during low solar activity conditions. 

To differentiate between the terrestrial and interstellar D$^-$ ions, \citet{rodriguez_etal:13a} split the available data into five intervals, presented in their Figures 5 and 6. Their boundaries correspond to DOYs 0, 15, 30, 45, 60, and 100. 
These intervals are marked in Figure~\ref{fig:spinD} with black vertical bars.
Deuterium ions identified as due to ISN D were found in interval 5, corresponding to DOYs 60---100.
This interval does not correspond to the highest flux.
However, the proportion of the counting rate from ISN D to that from the terrestrial D is the most favorable for detection of D. 

It is interesting to compare the IBEX-like case (blue dashed line at Figure~\ref{fig:spinD}) with the scenario proposed by \citet{kubiak_etal:23a}. During the first year, it is planned to observe the $\varepsilon$ lines 76\degr, 90\degr, and 104\degr. The proportion between the two counting rates for the proposed scenario of IMAP-Lo is more advantageous for detection than it was during the solar minimum for IBEX.
It is a little surprising because it could be intuitively expected that the joint action of radiation pressure and ionization rate would deplete ISN D more than ISN He. 
The absolute magnitude of the flux is indeed more advantageous during solar minimum conditions, but \cD/\cDterr{} turns out to be large during solar maximum conditions, for the IMAP-Lo geometry with $\varepsilon =104\degr$, and for much longer time, as illustrated in Figure \ref{fig:spinD}.

To make sure that the flux of ISN D in the region of high \cD/\cDterr{} for $\varepsilon = 104\degr$ is sufficient for potential detection of ISN D, we performed the following analysis.
We retrieved the simulated magnitude of ISN D flux for DOY 100, i.e., for the upper boundary of interval 5 defined by \citet{rodriguez_etal:13a} (Figure \ref{fig:spinD}, the blue broken line, IBEX-like case). 
This magnitude of the flux is plotted in Figure \ref{fig:Dflux} as the purple horizontal bar, and the corresponding DOY is marked in this figure by the left-side vertical bar.
With this, for $\varepsilon = 104\degr$, we checked for which DOY interval the simulated ISN D flux is larger than this value (Figure \ref{fig:spinD}, solid green line). 
The limit was found to be DOY 145. The prospective interval is marked with thick green in the figure.
This DOY, limiting the prospective time interval for ISN D observations, is marked with the right-side vertical bar in Figure \ref{fig:Dflux}.
In the found region, both the \cD/\cDterr{} ratio and the magnitude of the D flux are favorable for detection based on comparison with the findings in the IBEX-Lo data.

\subsection{Total counts}
\label{sec:totcount}
\noindent
Up to now, we have only discussed the maxima of the fluxes and spin angle-integrated counting rates, to illustrate the effects qualitatively. However, we are also interested in the total number of potentially detected atoms to assess statistical significance of the expected results, and to that end, we need to analyze the expected counting rates as a function of spin angle.
In Figure \ref{fig:spinD1}, we show as an example the fluxes (left panel) and the counting rates \cD{} and \cDterr{} (right panel) for DOY 121, $\varepsilon = 104\degr$, where the ratio $\cD/\cDterr$ is the greatest. 
Clearly, even though the ratio of the fluxes D/He is on the order of $10^{-4}$, the ratio of the counting rates can be larger than 1.

\begin{figure}[!ht]
\centering
\includegraphics[width=0.491\textwidth]{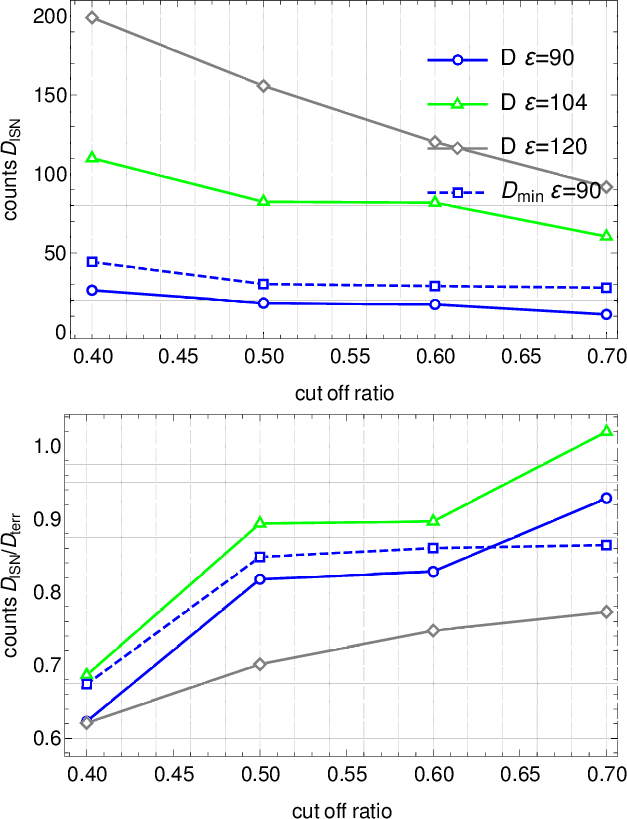}
\caption{
Upper panel: total number of counts of D$^-$ ions converted from ISN D atoms impacting the IMAP-Lo instrument (solid lines) for $\varepsilon = 90\degr, 104\degr$, and 120\degr{} (colors), for solar maximum conditions, shown as a function of the cutoff ratio, defined in Figure \ref{fig:spinD}. 
The broken line shows the equivalent count numbers simulated for the IBEX-Like viewing geometry ($\varepsilon = 90\degr$) for the solar minimum conditions. 
Lower panel: simulated ratios of \cD/\cDterr{} counts for various cutoff values; the colors and line styles equivalent to those used in the upper panel. 
}
\label{fig:countsRatioD}
\end{figure}

 
\begin{deluxetable*}{|lllllllll|}[!h]
\tablecaption{\label{tab:Dratio} Counts of the terrestrial and ISN D$^-$ ions integrated over various time intervals}
\tablehead{nr & sol. phase & elong & cut off ratio & counts D$_{\text{ISN}}$ & counts D$_{\text{terr}}$ & counts D$_{\text{ISN}}$/D$_{\text{terr                   }}$   & DOY start & DOY end }
\startdata
\hline
1 & sol. min  & 90 & 0.4 & 44.2877 & 65.7141 & 0.673946 & 77 & 121\\
 & & & 0.5 & 30.2688 & 35.6905 & 0.848092 & 81 & 117\\
 & & & 0.6 & 29.0039 & 33.7149 & 0.86027 & 81 & 109\\
 & & & 0.7 & 27.9371 & 32.3173 & 0.864461 & 81 & 105\\
\hline
2 & sol. max  & 90 & 0.4 & 26.3412 & 42.2738 & 0.62311 & 77 & 121\\
  & & & 0.5 & 18.247 & 22.3075 & 0.817975 & 81 & 117\\
  & & & 0.6 & 17.4883 & 21.1185 & 0.828103 & 81 & 109\\
  & & & 0.7 & 11.1203 & 11.9727 & 0.928806 & 85 & 105\\
\hline
 3 & sol. max  & 104 & 0.4 & 109.987 & 160.197 & 0.68657 & 105 & 185\\
  & & & 0.5 & 82.4068 & 92.1664 & 0.894109 & 109 & 177\\
  & & & 0.6 & 81.8384 & 91.2431 & 0.896927 & 109 & 169\\
  & & & 0.7 & 60.4352 & 59.2699 & 1.01966 & 113 & 165\\
  & & & 1.0 & 39.8675 & 36.7061 & 1.08613 & 117 & 145 \\
\hline
 4 & sol. max  & 120 & 0.4 & 199.088 & 320.651 & 0.620886 & 141 & 233\\
  & & & 0.5 & 155.794 & 222.1 & 0.701458 & 145 & 225\\
  & & & 0.6 & 120.234 & 160.854 & 0.747473 & 149 & 213\\
  & & & 0.7 & 91.7791 & 118.717 & 0.773094 & 153 & 205\\
\enddata
\end{deluxetable*}

In the measurements, we expect to see a certain ratio of D$^-$/H$^-$ events. 
This ratio is expected to be equal to $\xi_\text{D,terr} (1 + \cD/\cDterr)$, where $\xi_\text{D, terr}$ is the terrestrial abundance of D relative to H. 
For the times when $\cD \ll \cDterr$, the ratio will be equal to $\xi_\text{D,terr}$.
A signature of detection of ISN D is a statistically significant change in the observed D$^-$/H$^-$ ratio. 
This signature was used by \citet{rodriguez_etal:13a}.
We analyze the ratio \cD/\cDterr, which is the only variable term in the formula above.
We are interested in such observation conditions that this ratio is statistically significantly greater than 0, given the inevitable statistical scatter of the count numbers.
We compare the absolute count numbers for ISN D with the magnitude of the ratio of D$^-$counts due to ISN D to D$^-$ counts due to ISN He. This comparison is shown in Figure \ref{fig:countsRatioD}. 

In Table \ref{tab:Dratio}, we show estimates of the expected numbers of counts of the terrestrial and interstellar D$^-$ ions integrated over spin-angle range (55\degr, 115\degr) and the time intervals presented in the last two columns of the table.
These numbers, further on denoted as D$_\text{ISN}$ and D$_\text{terr}$, are presented for various cutoff levels of the ratio \cD/\cDterr{} presented in Figure \ref{fig:spinD}. 
These quantities were obtained assuming 100\% of usable observation time and are presented for the elongation angles 90\degr, 104\degr, and 120\degr{} for the solar maximum conditions. 
Additionally, for comparison with the IBEX observations conditions, we show them for the solar minimum conditions for $\varepsilon = 90\degr$.
The count numbers are obtained from integration over the time when the count rate \cD/\cDterr{} is larger than chosen cut off ratio.

The relative uncertainty of the ratio of counts D$_\text{ISN}$/D$_\text{terr}$ is given by $\sqrt{\text{(D}_\text{ISN})^{-1}+\text{(D}_\text{terr})^{-1}}$. For the count numbers as these listed in Table \ref{tab:Dratio}, the statistical uncertainties of the D$_\text{ISN}$/D$_\text{terr}$ ratio vary between $\sim 12$\% and $\sim 25\%$. 
Thus, the ratios listed in the table vary from 0 at a level of at least 2, and most often 3 $\sigma$, which offers a high probability of successful detection of ISN D during one year of IMAP operations.

However, this conclusion hinges on the actual percentage of usable data. 
On IBEX-Lo, backgrounds of all kinds resulted exposure times in three energy channels on the order of 10 hours per ISN observation season of 6 weeks\citep{swaczyna_etal:22b}.
If the exposure time length on IMAP is 10\% of the potentially available time, then the count numbers presented in Table \ref{tab:Dratio} are reduced tenfold and statistical uncertainties of the ratio  D$_\text{ISN}$/D$_\text{terr}$ are increased by a factor $\sqrt{10} \simeq 3.2$. 
With this, successful detection of ISN D in the data from one year of operations would require combining analysis of data from all three lines in the (DOY, $\varepsilon$) space.
Most likely, detection would be at a level of $\sim 1 \sigma$. 
Therefore, it is important to maximize the usable observation time for IMAP-Lo and to have a good grasp on the backgrounds.

Summing up this section, there is a realistic chance to measure ISN D during the nominal time of IMAP operations. 
The most advantageous geometry for this purpose is elongation $\varepsilon = 104\degr$. 
Despite measurements performed during high solar activity, the flux of ISN D is then comparable to the flux at IBEX during the solar minimum in 2009. 
The ratio \cD/\cDterr{} is not only larger than for the IBEX-like geometry during solar minimum, but also the time interval when this prerequisite is fulfilled is longer.
Certainly, however, detection of ISN D will require a meticulous analysis performed using a methodology similar to that used by \citet{rodriguez_etal:13a} and \citet{rodriguez_etal:14a}.

\section{Summary and conclusions}
\label{sec:conc}
\noindent
In this paper, we discuss opportunities for investigation of ISN H and D based on observations of the forthcoming NASA mission Interstellar Mapping and Acceleration Probe (IMAP). 
These studies are facilitated owing to the unique capability of the IMAP-Lo neutral atom camera to adjust the elongation angle $\varepsilon$ of its boresight to the Sun-oriented spin axis of the spacecraft.
We discuss these opportunities based on an extensive set of simulations of the fluxes of interstellar species.
We show that adopting a scenario for adjusting the elongation angle such that a set of parallel lines in the (DOY, $\varepsilon$) space is formed, it is possible to detect ISN D at a statistically significant level during one year of operation even during high-activity phase of the solar cycle, and to investigate various aspects of ISN H.

Adopting the scenario for adjustments of the $\varepsilon$ angle during two years of IMAP operations suggested by \citet{kubiak_etal:23a}, we identified the DOYs and elongations for which the observed flux of ISN H will have a small or negligible contribution from ISN He.
This greatly facilitates studies of ISN H. 
With this, we point out opportunities to observe the secondary population of ISN H with negligible or small contribution from the primary population.
We point out the opportunities to resolve the primary and secondary populations of ISN H in the approximation of two independent Maxwell-Boltzmann distribution functions at the entrance to the heliosphere.
We investigate signatures and point out locations in the (DOY, $\varepsilon$) space to look for the departure of the flow of ISN H observed by IMAP-Lo from a model with a single Maxwell-Boltzmann population with the flow parameters modified within the outer heliosheath. 

We also identify the observation conditions for which the observed signal can be expected to be especially sensitive to the flow parameters of the primary and secondary populations, constrained by insight from existing analyses, which suggest that the mean flow parameters of ISN H inside the heliosphere agree with those inferred by \citet{lallement_etal:05a} from spectral observations of the helioglow. 
Finally, we point out the sensitivity of the observed ISN H flux to details of radiation pressure, in particular to its dependence on radial speed of ISN H atoms. 
We find that this dependence strongly affects the observed signal, so a good understanding of the evolution of the solar Lyman-$\alpha$ profile is necessary for successful analysis of observations of ISN H on IMAP-Lo.

IMAP-Lo will address key questions of the heliospheric physics, and even general astrophysics.
Following the scenario of adjustment of the instrument boresight suggested by \citet{kubiak_etal:23a} during two years of IMAP nominal mission we have shown how -- with appropriate modeling support -- measurements are used to understand the interaction of the unperturbed ISN H with the plasma in the outer heliosheath and to establish the abundance of ISN D relative to ISN H at the termination shock of the solar wind. 
Maintaining parallel observations with the IBEX mission will likely provide a better understanding of both the existing IBEX data and future IMAP data, which during the IMAP mission will span almost two full cycles of solar activity.

\section*{Acknowledgments}
The authors wish to dedicate this paper to the memory of Professor Vladimir B. Baranov (1934---2023), one of the pioneers of heliospheric and interstellar neutral atom studies, a stern tutor, a warm man, and a sincere, good friend. He will be remembered. 

We are obliged to Pawe{\l} Swaczyna for helpful discussions.

The work at CBK PAN was supported by the National Science Centre, Poland (grant 2019/35/B/ST9/01241).
\bibliographystyle{aasjournal}
\bibliography{iplbib}

\begin{thebibliography}{}
\expandafter\ifx\csname natexlab\endcsname\relax\def\natexlab#1{#1}\fi

\bibitem[{{Baranov} \& {Izmodenov}(2006)}]{baranov_izmodenov:06a}
{Baranov}, V.~B., \& {Izmodenov}, V.~V. 2006, Fluid Dynamics, 41, 689

\bibitem[{Baranov {et~al.}(1998)Baranov, Izmodenov, \&
  Malama}]{baranov_etal:98a}
Baranov, V.~B., Izmodenov, V.~V., \& Malama, Y.~G. 1998, \jgr, 103, 9575

\bibitem[{Baranov \& Malama(1993)}]{baranov_malama:93}
Baranov, V.~B., \& Malama, Y.~G. 1993, \jgr, 98, 15157

\bibitem[{B{\"o}hlke {et~al.}(2005)B{\"o}hlke, de~Laeter, {DeBi{\`e}vre},
  Hikada, Peiser, Rosman, \& Taylor}]{boehlke_etal:05a}
B{\"o}hlke, J.~K., de~Laeter, J.~R., {DeBi{\`e}vre}, P., {et~al.} 2005, J.
  Phys. Chem Ref. Data, 34, 51

\bibitem[{Bzowski {et~al.}(1997)Bzowski, Fahr, Ruci{\'n}ski, \&
  Scherer}]{bzowski_etal:97}
Bzowski, M., Fahr, H.~J., Ruci{\'n}ski, D., \& Scherer, H. 1997, \aap, 326, 396

\bibitem[{{Bzowski} {et~al.}(2017){Bzowski}, {Kubiak}, {Czechowski}, \&
  {Grygorczuk}}]{bzowski_etal:17a}
{Bzowski}, M., {Kubiak}, M.~A., {Czechowski}, A., \& {Grygorczuk}, J. 2017,
  \apj, 845, 15

\bibitem[{Bzowski {et~al.}(2022)Bzowski, Kubiak, M{\"o}bius, \&
  Schwadron}]{bzowski_etal:22a}
Bzowski, M., Kubiak, M.~A., M{\"o}bius, E., \& Schwadron, N.~A. 2022, \apj,
  938, 148

\bibitem[{{Bzowski} {et~al.}(2023){Bzowski}, {Kubiak}, {M{\"o}bius}, \&
  {Schwadron}}]{bzowski_etal:23a}
{Bzowski}, M., {Kubiak}, M.~A., {M{\"o}bius}, E., \& {Schwadron}, N.~A. 2023,
  \apjs, 265, 24

\bibitem[{Bzowski {et~al.}(2023)Bzowski, Kubiak, Strumik, Kowalska-Leszczynska,
  Porowski, \& Qu{\'e}merais}]{bzowski_etal:23b}
Bzowski, M., Kubiak, M.~A., Strumik, M., {et~al.} 2023, \apj, 952, 2

\bibitem[{{Bzowski} {et~al.}(2008){Bzowski}, {M{\"o}bius}, {Tarnopolski},
  {Izmodenov}, \& {Gloeckler}}]{bzowski_etal:08a}
{Bzowski}, M., {M{\"o}bius}, E., {Tarnopolski}, S., {Izmodenov}, V., \&
  {Gloeckler}, G. 2008, \aap, 491, 7

\bibitem[{{Bzowski} {et~al.}(2009){Bzowski}, {M{\"o}bius}, {Tarnopolski},
  {Izmodenov}, \& {Gloeckler}}]{bzowski_etal:09a}
---. 2009, \ssr, 143, 177

\bibitem[{Bzowski {et~al.}(2012)Bzowski, Kubiak, M{\"o}bius, Bochsler, Leonard,
  Heirtzler, Kucharek, Sok{\'{o}}{\l}, H{\l}ond, Crew, Schwadron, Fuselier, \&
  McComas}]{bzowski_etal:12a}
Bzowski, M., Kubiak, M.~A., M{\"o}bius, E., {et~al.} 2012, \apjs, 198, 12

\bibitem[{{Bzowski} {et~al.}(2015){Bzowski}, {Swaczyna}, {Kubiak},
  {Sok\'{o}{\l}}, {Fuselier}, {Galli}, {Heirtzler}, {Kucharek}, {Leonard},
  {McComas}, {M{\"o}bius}, {Schwadron}, \& {Wurz}}]{bzowski_etal:15a}
{Bzowski}, M., {Swaczyna}, P., {Kubiak}, M.~A., {et~al.} 2015, \apjs, 220, 28

\bibitem[{Bzowski {et~al.}(2019)Bzowski, Czechowski, Frisch, Fuselier, Galli,
  Grygorczuk, Heerikhuisen, Kubiak, Kucharek, McComas, M{\"o}bius, Schwadron,
  Slavin, Sok{\'o}{\l}, Swaczyna, Wurz, \& Zirnstein}]{bzowski_etal:19a}
Bzowski, M., Czechowski, A., Frisch, P., {et~al.} 2019, \apj, 882, 60

\bibitem[{{Fuselier} {et~al.}(2009){Fuselier}, {Bochsler}, {Chornay}, {Clark},
  {Crew}, {Dunn}, {Ellis}, {Friedmann}, {Funsten}, {Ghielmetti}, {Googins},
  {Granoff}, {Hamilton}, {Hanley}, {Heirtzler}, {Hertzberg}, {Isaac}, {King},
  {Knauss}, {Kucharek}, {Kudirka}, {Livi}, {Lobell}, {Longworth}, {Mashburn},
  {McComas}, {M{\"o}bius}, {Moore}, {Moore}, {Nemanich}, {Nolin}, {O'Neal},
  {Piazza}, {Peterson}, {Pope}, {Rosmarynowski}, {Saul}, {Scherrer}, {Scheer},
  {Schlemm}, {Schwadron}, {Tillier}, {Turco}, {Tyler}, {Vosbury}, {Wieser},
  {Wurz}, \& {Zaffke}}]{fuselier_etal:09b}
{Fuselier}, S.~A., {Bochsler}, P., {Chornay}, D., {et~al.} 2009, \ssr, 146, 117

\bibitem[{{Galli} {et~al.}(2019){Galli}, {Wurz}, {Rahmanifard}, {M{\"o}bius},
  {Schwadron}, {Kucharek}, {Heirtzler}, {Fairchild}, {Bzowski}, {Kubiak},
  {Kowalska-Leszczy{\'n}ska}, {Sok{\'o}{\l}}, {Fuselier}, {Swaczyna}, \&
  {McComas}}]{galli_etal:19a}
{Galli}, A., {Wurz}, P., {Rahmanifard}, F., {et~al.} 2019, \apj, 871, 52

\bibitem[{{H{\'e}brard} {et~al.}(1999){H{\'e}brard}, {Mallouris}, {Ferlet},
  {Koester}, {Lemoine}, {Vidal-Madjar}, \& {York}}]{hebrard_etal:99a}
{H{\'e}brard}, G., {Mallouris}, C., {Ferlet}, R., {et~al.} 1999, \aap, 350, 643

\bibitem[{{Heerikhuisen} \& {Pogorelov}(2010)}]{heerikhuisen_pogorelov:10a}
{Heerikhuisen}, J., \& {Pogorelov}, N.~V. 2010, in Astronomical Society of the
  Pacific Conference Series, Vol. 429, Numerical Modeling of Space Plasma
  Flows, Astronum-2009, ed. N.~V. {Pogorelov}, E.~{Audit}, \& G.~P. {Zank},
  227--232

\bibitem[{{Izmodenov} {et~al.}(2003){Izmodenov}, {Gloeckler}, \&
  {Malama}}]{izmodenov_etal:03b}
{Izmodenov}, V., {Gloeckler}, G., \& {Malama}, Y. 2003, \grl, 30, 3

\bibitem[{Izmodenov(2001)}]{izmodenov:01}
Izmodenov, V.~V. 2001, \ssr, 97, 385

\bibitem[{{Katushkina} {et~al.}(2021){Katushkina}, {Galli}, {Izmodenov}, \&
  {Alexashov}}]{katushkina_etal:21b}
{Katushkina}, O.~A., {Galli}, A., {Izmodenov}, V.~V., \& {Alexashov}, D.~B.
  2021, \mnras, 501, 1633

\bibitem[{{Katushkina} \& {Izmodenov}(2010)}]{katushkina_izmodenov:10}
{Katushkina}, O.~A., \& {Izmodenov}, V.~V. 2010, Astronomy Letters, 36, 297

\bibitem[{{Katushkina} {et~al.}(2015){Katushkina}, {Izmodenov}, \&
  {Alexashov}}]{katushkina_etal:15a}
{Katushkina}, O.~A., {Izmodenov}, V.~V., \& {Alexashov}, D.~B. 2015, \mnras,
  446, 2929

\bibitem[{{Koutroumpa} {et~al.}(2017){Koutroumpa}, {Qu{\'e}merais},
  {Katushkina}, {Lallement}, {Bertaux}, \& {Schmidt}}]{koutroumpa_etal:17a}
{Koutroumpa}, D., {Qu{\'e}merais}, E., {Katushkina}, O., {et~al.} 2017, \aap,
  598, A12

\bibitem[{{Kowalska-Leszczynska} {et~al.}(2020){Kowalska-Leszczynska},
  {Bzowski}, {Kubiak}, \& {Sok{\'o}{\l}}}]{IKL:20a}
{Kowalska-Leszczynska}, I., {Bzowski}, M., {Kubiak}, M.~A., \& {Sok{\'o}{\l}},
  J.~M. 2020, \apjs, 247, 62

\bibitem[{{Kowalska-Leszczynska}
  {et~al.}(2018{\natexlab{a}}){Kowalska-Leszczynska}, {Bzowski},
  {Sok{\'o}{\l}}, \& {Kubiak}}]{IKL:18b}
{Kowalska-Leszczynska}, I., {Bzowski}, M., {Sok{\'o}{\l}}, J.~M., \& {Kubiak},
  M.~A. 2018{\natexlab{a}}, \apj, 868, 49

\bibitem[{{Kowalska-Leszczynska}
  {et~al.}(2018{\natexlab{b}}){Kowalska-Leszczynska}, {Bzowski},
  {Sok{\'o}{\l}}, \& {Kubiak}}]{IKL:18a}
---. 2018{\natexlab{b}}, \apj, 852, 15

\bibitem[{{Kowalska-Leszczynska} {et~al.}(2022){Kowalska-Leszczynska},
  {Kubiak}, \& {Bzowski}}]{IKL:22a}
{Kowalska-Leszczynska}, I., {Kubiak}, M.~A., \& {Bzowski}, M. 2022, \apj, 926,
  27

\bibitem[{{Kowalska-Leszczynska} {et~al.}(2023){Kowalska-Leszczynska},
  {Kubiak}, \& {Bzowski}}]{IKL:23a}
---. 2023, \apj, 950, 98

\bibitem[{{Kretzschmar} {et~al.}(2018){Kretzschmar}, {Snow}, \&
  {Curdt}}]{kretzschmar_etal:18a}
{Kretzschmar}, M., {Snow}, M., \& {Curdt}, W. 2018, \grl, 45, 2138

\bibitem[{Kubiak {et~al.}(2021)Kubiak, Bzowski, Kowalska-Leszczynska, \&
  Strumik}]{kubiak_etal:21b}
Kubiak, M.~A., Bzowski, M., Kowalska-Leszczynska, I., \& Strumik, M. 2021,
  \apjs, 254, 17

\bibitem[{Kubiak {et~al.}(2019)Kubiak, Bzowski, \&
  Sok{\'o}{\l}}]{kubiak_etal:19a}
Kubiak, M.~A., Bzowski, M., \& Sok{\'o}{\l}, J. 2019, \apj, 882, 114

\bibitem[{Kubiak {et~al.}(2013)Kubiak, Bzowski, Sok{\'o\l}, M{\"o}bius,
  Rodr{\'i}guez, Wurz, \& McComas}]{kubiak_etal:13a}
Kubiak, M.~A., Bzowski, M., Sok{\'o\l}, J.~M., {et~al.} 2013, \aap, 556, A39

\bibitem[{Kubiak {et~al.}(2023)Kubiak, Bzowski, Swaczyna, M{\"o}bius,
  Schwadron, \& McComas}]{kubiak_etal:23a}
Kubiak, M.~A., Bzowski, M., Swaczyna, P., {et~al.} 2023, \apj, 269, 23

\bibitem[{{Kubiak} {et~al.}(2014){Kubiak}, {Bzowski}, {Sok{\'o}{\l}},
  {Swaczyna}, {Grzedzielski}, {Alexashov}, {Izmodenov}, {Moebius}, {Leonard},
  {Fuselier}, {Wurz}, \& {McComas}}]{kubiak_etal:14a}
{Kubiak}, M.~A., {Bzowski}, M., {Sok{\'o}{\l}}, J.~M., {et~al.} 2014, \apjs,
  213, 29

\bibitem[{Kubiak {et~al.}(2016)Kubiak, Swaczyna, Bzowski, Sok{\'o}{\l},
  Fuselier, Galli, Heirtzler, Kucharek, Leonard, McComas, Park, Schwadron, \&
  Wurz}]{kubiak_etal:16a}
Kubiak, M.~A., Swaczyna, P., Bzowski, M., {et~al.} 2016, \apjs, 223, 35

\bibitem[{Lallement \& Bertaux(1990)}]{lallement_bertaux:90a}
Lallement, R., \& Bertaux, J.~L. 1990, \aap, 231, L3

\bibitem[{{Lallement} {et~al.}(1993){Lallement}, {Bertaux}, \&
  {Clarke}}]{lallement_etal:93a}
{Lallement}, R., {Bertaux}, J.-L., \& {Clarke}, J.~T. 1993, Science, 260, 1095

\bibitem[{{Lallement} {et~al.}(2005){Lallement}, {Qu{\' e}merais}, {Bertaux},
  {Ferron}, {Koutroumpa}, \& {Pellinen}}]{lallement_etal:05a}
{Lallement}, R., {Qu{\' e}merais}, E., {Bertaux}, J.~L., {et~al.} 2005,
  Science, 307, 1447

\bibitem[{{Lallement} {et~al.}(2010{\natexlab{a}}){Lallement}, {Qu{\'e}merais},
  {Koutroumpa}, {Bertaux}, {Ferron}, {Schmidt}, \& {Lamy}}]{lallement_etal:10a}
{Lallement}, R., {Qu{\'e}merais}, E., {Koutroumpa}, D., {et~al.}
  2010{\natexlab{a}}, Twelfth International Solar Wind Conference, 1216, 555

\bibitem[{{Lallement} {et~al.}(2010{\natexlab{b}}){Lallement}, {Qu{\'e}merais},
  {Lamy}, {Bertaux}, {Ferron}, \& {Schmidt}}]{lallement_etal:10b}
{Lallement}, R., {Qu{\'e}merais}, E., {Lamy}, P., {et~al.} 2010{\natexlab{b}},
  in Astronomical Society of the Pacific Conference Series, Vol. 428, SOHO-23:
  Understanding a Peculiar Solar Minimum, ed. {S.~R.~Cranmer, J.~T.~Hoeksema,
  \& J.~L.~Kohl}, 253--258

\bibitem[{Lemaire {et~al.}(2015)Lemaire, Vial, Curdt, Sch{\"u}hle, \&
  Wilhelm}]{lemaire_etal:15a}
Lemaire, P., Vial, J., Curdt, W., Sch{\"u}hle, U., \& Wilhelm, K. 2015, \aap,
  581, A26

\bibitem[{{Linsky} {et~al.}(2006){Linsky}, {Draine}, {Moos}, {Jenkins}, {Wood},
  {Oliveira}, {Blair}, {Friedman}, {Gry}, {Knauth}, {Kruk}, {Lacour}, {Lehner},
  {Redfield}, {Shull}, {Sonneborn}, \& {Williger}}]{linsky_etal:06a}
{Linsky}, J.~L., {Draine}, B.~T., {Moos}, H.~W., {et~al.} 2006, \apj, 647, 1106

\bibitem[{Machol {et~al.}(2019)Machol, Snow, Woodraska, Woods, Viereck, \&
  Coddington}]{machol_etal:19a}
Machol, J.~L., Snow, M., Woodraska, D., {et~al.} 2019, Earth and Space Science,
  6, 2263

\bibitem[{McComas {et~al.}(2018)McComas, Christian, Schwadron, Fox, Westlake,
  Allegrini, Baker, Biesecker, Bzowski, Clark, Cohen, Cohen, Dayeh, Decker,
  de~Nolfo, Desai, andH.A. Elliott, Fahr, Frisch, Funsten, Fuselier, Galli,
  Galvin, Giacalone, Gkioulidou, Guo, Horanyi, Isenberg, Janzen, Kistler,
  Korreck, Kubiak, Kucharek, Larsen, Leske, Lugaz, Luhmann, Matthaeus, Mitchel,
  Moebius, Ogasawara, Reisenfeld, Richardson, Russell, Sok{\'o}{\l}, Spence,
  Skoug, Sternovsky, Swaczyna, Szalay, Tokumaru, andP. Wurz, Zank, \&
  Zirnstein}]{mccomas_etal:18b}
McComas, D., Christian, E., Schwadron, N., {et~al.} 2018, \ssr, 214, 116

\bibitem[{{McComas} {et~al.}(2009){McComas}, {Allegrini}, {Bochsler},
  {Bzowski}, {Christian}, {Crew}, {DeMajistre}, {Fahr}, {Fichtner}, {Frisch},
  {Funsten}, {Fuselier}, {Gloeckler}, {Gruntman}, {Heerikhuisen}, {Izmodenov},
  {Janzen}, {Knappenberger}, {Krimigis}, {Kucharek}, {Lee}, {Livadiotis},
  {Livi}, {MacDowall}, {Mitchell}, {M{\"o}bius}, {Moore}, {Pogorelov},
  {Reisenfeld}, {Roelof}, {Saul}, {Schwadron}, {Valek}, {Vanderspek}, {Wurz},
  \& {Zank}}]{mccomas_etal:09c}
{McComas}, D.~J., {Allegrini}, F., {Bochsler}, P., {et~al.} 2009, Science, 326,
  959

\bibitem[{McComas {et~al.}(2020)McComas, Bzowski, Dayeh, DeMajistre, Funsten,
  Janzen, Kowalska-Leszczy{\'{n}}ska, Kubiak, Schwadron, Sok{\'{o}}{\l},
  Szalay, Tokumaru, \& Zirnstein}]{mccomas_etal:20a}
McComas, D.~J., Bzowski, M., Dayeh, M.~A., {et~al.} 2020, \apjs, 248, 26

\bibitem[{{M{\"o}bius} {et~al.}(2009{\natexlab{a}}){M{\"o}bius}, {Kucharek},
  {Clark}, {O'Neill}, {Petersen}, {Bzowski}, {Saul}, {Wurz}, {Fuselier},
  {Izmodenov}, {McComas}, {M{\"u}ller}, \& {Alexashov}}]{mobius_etal:09a}
{M{\"o}bius}, E., {Kucharek}, H., {Clark}, G., {et~al.} 2009{\natexlab{a}},
  \ssr, 146, 149

\bibitem[{{M{\"o}bius} {et~al.}(2009{\natexlab{b}}){M{\"o}bius}, {Bochsler},
  {Bzowski}, {Crew}, {Funsten}, {Fuselier}, {Ghielmetti}, {Heirtzler},
  {Izmodenov}, {Kubiak}, {Kucharek}, {Lee}, {Leonard}, {McComas}, {Petersen},
  {Saul}, {Scheer}, {Schwadron}, {Witte}, \& {Wurz}}]{mobius_etal:09b}
{M{\"o}bius}, E., {Bochsler}, P., {Bzowski}, M., {et~al.} 2009{\natexlab{b}},
  Science, 326, 969

\bibitem[{Park {et~al.}(2016)Park, Kucharek, M{\"o}bius, Galli, Kubiak,
  Bzowski, \& McComas}]{park_etal:16a}
Park, J., Kucharek, H., M{\"o}bius, E., {et~al.} 2016, \apj, 833, 130

\bibitem[{Porowski {et~al.}(2022)Porowski, Bzowski, \&
  Tokumaru}]{porowski_etal:22a}
Porowski, C., Bzowski, M., \& Tokumaru, M. 2022, \apjs, 259, 2

\bibitem[{Porowski {et~al.}(2023)Porowski, Bzowski, \&
  Tokumaru}]{porowski_etal:23a}
---. 2023, \apjs, 264, 11

\bibitem[{{Prantzos}(1996)}]{prantzos:96a}
{Prantzos}, N. 1996, \aap, 310, 106

\bibitem[{{Prantzos}(2007)}]{prantzos:07a}
---. 2007, \ssr, 130, 27

\bibitem[{Qu{\'e}merais {et~al.}(1999)Qu{\'e}merais, Bertaux, Lallement,
  Berth{\'e}, Kyr{\"o}l{\"a}, \& Schmidt}]{quemerais_etal:99}
Qu{\'e}merais, E., Bertaux, J.-L., Lallement, R., {et~al.} 1999, \jgr, 104,
  12585

\bibitem[{Rahmanifard {et~al.}(2019)Rahmanifard, M{\"o}bius, Schwadron, Galli,
  Richards, Kucharek, Sok{\'{o}}{\l}, Heirtzler, Lee, Bzowski,
  Kowalska-Leszczynska, Kubiak, Wurz, Fuselier, \&
  McComas}]{rahmanifard_etal:19a}
Rahmanifard, F., M{\"o}bius, E., Schwadron, N.~A., {et~al.} 2019, \apj, 887,
  217

\bibitem[{{Rahmanifard} {et~al.}(2024){Rahmanifard}, {Swaczyna}, {Zirnstein},
  {Heerikhuisen}, {Galli}, {Sok{\'o}{\l}}, {Schwadron}, {M{\"o}bius},
  {McComas}, \& {Fuselier}}]{rahmanifard_etal:23a}
{Rahmanifard}, F., {Swaczyna}, P., {Zirnstein}, E.~J., {et~al.} 2024, \apj,
  959, 129

\bibitem[{{Ratkiewicz} {et~al.}(2002){Ratkiewicz}, {Barnes}, {M{\"u}ller},
  {Zank}, \& {Webb}}]{ratkiewicz_etal:02a}
{Ratkiewicz}, R., {Barnes}, A., {M{\"u}ller}, H.-R., {Zank}, G.~P., \& {Webb},
  G.~M. 2002, \asr, 29, 433

\bibitem[{Reisenfeld {et~al.}(2021)Reisenfeld, Bzowski, Funsten, Heerikhuisen,
  Janzen, Kubiak, McComas, Schwadron, Sok{\'o}{\l}, Zimorino, \&
  Zirnstein}]{reisenfeld_etal:21a}
Reisenfeld, D.~B., Bzowski, M., Funsten, H., {et~al.} 2021, \apjs, 254, 40

\bibitem[{{Rodr{\'i}guez Moreno} {et~al.}(2014){Rodr{\'i}guez Moreno}, {Wurz},
  {Saul}, {Bzowski}, {Kubiak}, {Sok{\'o}{\l}}, {Frisch}, {Fuselier}, {McComas},
  {M{\"o}bius}, \& {Schwadron}}]{rodriguez_etal:14a}
{Rodr{\'i}guez Moreno}, D., {Wurz}, P., {Saul}, L., {et~al.} 2014, Entropy, 16,
  1134

\bibitem[{{Rodr{\'i}guez Moreno} {et~al.}(2013){Rodr{\'i}guez Moreno}, Wurz,
  Saul, Bzowski, Kubiak, Sok{\'o\l}, Frisch, Fuselier, McComas, M{\"o}bius, \&
  Schwadron}]{rodriguez_etal:13a}
{Rodr{\'i}guez Moreno}, D.~F., Wurz, P., Saul, L., {et~al.} 2013, \aap, 557,
  A125

\bibitem[{{Saul} {et~al.}(2012){Saul}, {Wurz}, {Rodriguez}, {Scheer},
  {M{\"o}bius}, {Schwadron}, {Kucharek}, {Leonard}, {Bzowski}, {Fuselier},
  {Crew}, \& {McComas}}]{saul_etal:12a}
{Saul}, L., {Wurz}, P., {Rodriguez}, D., {et~al.} 2012, \apjs, 198, 14

\bibitem[{Saul {et~al.}(2013)Saul, Bzowski, Fuselier, Kubiak, McComas,
  M{\"o}bius, Sok{'o}{\l}, Rodr{\'i}guez, Scheer, \& Wurz}]{saul_etal:13a}
Saul, L., Bzowski, M., Fuselier, S., {et~al.} 2013, \apj, 767, 130

\bibitem[{{Schwadron} {et~al.}(2015){Schwadron}, {M{\"o}bius}, {Leonard},
  {Fuselier}, {Bzowski}, {Frisch}, {Heirtzler}, {Kubiak}, {Kucharek}, {Lee},
  {McComas}, {Rahmanifard}, {Sok\'{o}{\l}}, \& {Swaczyna}}]{schwadron_etal:15a}
{Schwadron}, N., {M{\"o}bius}, E., {Leonard}, T., {et~al.} 2015, \apjs, 220, 25

\bibitem[{{Schwadron} \& {McComas}(2021)}]{schwadron_mccomas:21a}
{Schwadron}, N.~A., \& {McComas}, D.~J. 2021, \apj, 914, 129

\bibitem[{{Schwadron} {et~al.}(2013){Schwadron}, {Moebius}, {Kucharek}, {Lee},
  {French}, {Saul}, {Wurz}, {Bzowski}, {Fuselier}, {Livadiotis}, {McComas},
  {Frisch}, {Gruntman}, \& {Mueller}}]{schwadron_etal:13a}
{Schwadron}, N.~A., {Moebius}, E., {Kucharek}, H., {et~al.} 2013, \apj, 775, 86

\bibitem[{{Schwadron} {et~al.}(2014){Schwadron}, {Adams}, {Christian},
  {Desiati}, {Frisch}, {Funsten}, {Jokipii}, {McComas}, {Moebius}, \&
  {Zank}}]{schwadron_etal:14a}
{Schwadron}, N.~A., {Adams}, F.~C., {Christian}, E.~R., {et~al.} 2014, Science,
  343, 988

\bibitem[{Schwadron {et~al.}(2022)Schwadron, M{\"o}bius, McComas, Bower, Bower,
  Bzowski, Fuselier, Heirtzler, Kubiak, Lee, Rahmanifard, Sok{\'o}{\l},
  Swaczyn, \& Winslow}]{schwadron_etal:22a}
Schwadron, N.~A., M{\"o}bius, E., McComas, D.~J., {et~al.} 2022, \apjs, 258, 7

\bibitem[{{Sok{\'o}{\l}} {et~al.}(2019{\natexlab{a}}){Sok{\'o}{\l}}, {Bzowski},
  \& {Tokumaru}}]{sokol_etal:19a}
{Sok{\'o}{\l}}, J.~M., {Bzowski}, M., \& {Tokumaru}, M. 2019{\natexlab{a}},
  \apj, 872, 57

\bibitem[{{Sok{\'o}{\l}} {et~al.}(2019{\natexlab{b}}){Sok{\'o}{\l}}, {Kubiak},
  {Bzowski}, {M{\"o}bius}, \& {Schwadron}}]{sokol_etal:19c}
{Sok{\'o}{\l}}, J.~M., {Kubiak}, M.~A., {Bzowski}, M., {M{\"o}bius}, E., \&
  {Schwadron}, N. 2019{\natexlab{b}}, \apjs, 245, 28

\bibitem[{{Sok\'{o}{\l}} {et~al.}(2015){Sok\'{o}{\l}}, {Kubiak}, {Bzowski}, \&
  {Swaczyna}}]{sokol_etal:15b}
{Sok\'{o}{\l}}, J.~M., {Kubiak}, M.~A., {Bzowski}, M., \& {Swaczyna}, P. 2015,
  \apjs, 220, 27

\bibitem[{{Sok{\'o}{\l}} {et~al.}(2020){Sok{\'o}{\l}}, {McComas}, {Bzowski}, \&
  {Tokumaru}}]{sokol_etal:20a}
{Sok{\'o}{\l}}, J.~M., {McComas}, D.~J., {Bzowski}, M., \& {Tokumaru}, M. 2020,
  \apj, 897, 179

\bibitem[{Swaczyna {et~al.}(2021)Swaczyna, Rahmanifard, Zirnstein, McComas, \&
  Heerikhuisen}]{swaczyna_etal:21a}
Swaczyna, P., Rahmanifard, F., Zirnstein, E., McComas, D., \& Heerikhuisen, J.
  2021, \apjl, 911, L36

\bibitem[{{Swaczyna} {et~al.}(2023{\natexlab{a}}){Swaczyna}, {Rahmanifard},
  {Zirnstein}, \& {Heerikhuisen}}]{swaczyna_etal:23a}
{Swaczyna}, P., {Rahmanifard}, F., {Zirnstein}, E.~J., \& {Heerikhuisen}, J.
  2023{\natexlab{a}}, \apj, 943, 74

\bibitem[{Swaczyna {et~al.}(2020)Swaczyna, McComas, Zirnstein, Sok{\'o}{\l},
  Elliott, Bzowski, Kubiak, Richardson, Kowalska-Leszczynska, Stern, Weaver,
  Olkin, Singer, \& Spencer}]{swaczyna_etal:20a}
Swaczyna, P., McComas, D., Zirnstein, E.~J., {et~al.} 2020, \apj, 903, 48

\bibitem[{Swaczyna {et~al.}(2022)Swaczyna, Kubiak, Bzowski, Bower, Fuselier,
  Galli, Heirtzler, McComas, M{\"o}bius, Rahmanifard, \&
  Schwadron}]{swaczyna_etal:22b}
Swaczyna, P., Kubiak, M.~A., Bzowski, M., {et~al.} 2022, \apjs, 259, 42

\bibitem[{{Swaczyna} {et~al.}(2023{\natexlab{b}}){Swaczyna}, {Bzowski},
  {Heerikhuisen}, {Kubiak}, {Rahmanifard}, {Zirnstein}, {Fuselier}, {Galli},
  {McComas}, {M{\"o}bius}, \& {Schwadron}}]{swaczyna_etal:23c}
{Swaczyna}, P., {Bzowski}, M., {Heerikhuisen}, J., {et~al.} 2023{\natexlab{b}},
  \apj, 953, 107

\bibitem[{{Tarnopolski} \& {Bzowski}(2008)}]{tarnopolski_bzowski:08a}
{Tarnopolski}, S., \& {Bzowski}, M. 2008, \aap, 483, L35

\bibitem[{Tarnopolski \& Bzowski(2009)}]{tarnopolski_bzowski:09}
Tarnopolski, S., \& Bzowski, M. 2009, \aap, 493, 207

\bibitem[{{Witte}(2004)}]{witte:04}
{Witte}, M. 2004, \aap, 426, 835

\bibitem[{Wurz {et~al.}(2008)Wurz, Saul, Scheer, M{\"o}bius, Kucharek, \&
  Fuselier}]{wurz_etal:08a}
Wurz, P., Saul, L., Scheer, J.~A., {et~al.} 2008, J. Appl. Phys., 103, 054904

\bibitem[{{Zirnstein} {et~al.}(2016{\natexlab{a}}){Zirnstein}, {Funsten},
  {Heerikhuisen}, {McComas}, {Schwadron}, \& {Zank}}]{zirnstein_etal:16c}
{Zirnstein}, E., {Funsten}, H., {Heerikhuisen}, J., {et~al.}
  2016{\natexlab{a}}, \apj, 826, 58

\bibitem[{{Zirnstein} {et~al.}(2016{\natexlab{b}}){Zirnstein}, {Heerikhuisen},
  {Funsten}, {Livadiotis}, {McComas}, \& {Pogorelov}}]{zirnstein_etal:16b}
{Zirnstein}, E.~J., {Heerikhuisen}, J., {Funsten}, H.~O., {et~al.}
  2016{\natexlab{b}}, \apjl, 818, L18

\bibitem[{{Zirnstein} {et~al.}(2022){Zirnstein}, {Shrestha}, {McComas},
  {Dayeh}, {Heerikhuisen}, {Reisenfeld}, {Sok{\'o}{\l}}, \&
  {Swaczyna}}]{zirnstein_etal:22b}
{Zirnstein}, E.~J., {Shrestha}, B.~L., {McComas}, D.~J., {et~al.} 2022, Nature
  Astronomy, 6, 1398

\end{thebibliography}

\end{document}